\documentclass[11pt,a4paper]{article}

\usepackage{amsmath,amsfonts,graphicx,natbib,psfrag,color,pstool}
\usepackage{algorithm, algorithmic}
\usepackage[textsize=small]{todonotes}
\usepackage{booktabs}
 \usepackage{hyperref}

\newcommand{\LNA}{\textrm{LNA}}

\title{Accelerating inference for stochastic kinetic models}

\author{Tom E. Lowe$^1$\footnote{t.lowe@ncl.ac.uk}, Andrew   Golightly$^2$\footnote{andrew.golightly@durham.ac.uk}, Chris Sherlock$^3$\footnote{c.sherlock@lancaster.ac.uk}}
\date{\small $^1$ School of Mathematics, Statistics and Physics, Newcastle University, UK\\$^2$ Department of Mathematical Sciences, Durham University, UK\\$^3$ Department of Mathematics and Statistics, Lancaster University, UK}

\begin{document}
\maketitle
\begin{abstract}
Stochastic kinetic models (SKMs) are increasingly used to account for the inherent stochasticity exhibited by interacting populations of species in areas such as epidemiology, population ecology and systems biology. Species numbers are modelled using a  continuous-time stochastic process, and, depending on the 
application area of interest, this will typically take the form of a Markov jump process or an It\^o diffusion process. Widespread use of these models is typically precluded by their computational complexity. In particular, performing exact fully Bayesian inference in either modelling framework is challenging due to the intractability of the observed data likelihood, necessitating the use of computationally intensive techniques such as particle Markov chain Monte Carlo (particle MCMC). It is proposed to increase the computational and statistical efficiency of this approach by leveraging the tractability of an inexpensive surrogate derived directly from either the jump or diffusion process. The surrogate is used in three ways: in the design of a gradient-based parameter proposal, to construct an appropriate bridge and in the first stage of a delayed-acceptance step. The resulting approach, which exactly targets the posterior of interest, offers substantial gains in efficiency over a standard particle MCMC implementation.
\end{abstract}


\section{Introduction}
\label{sec:intro}

A stochastic kinetic model (SKM) typically refers to a reaction network, an associated rate law and a probabilistic description of the reaction dynamics \citep[see e.g.][]{Wilkinson06}. A Markov jump process (MJP) provides the most natural description of the time-course behaviour of the species involved in the reaction network. In scenarios where species numbers can be reasonably regarded as continuous, it is commonplace to approximate the MJP to give an It\^o stochastic differential equation known as the chemical Langevin equation (CLE). The modelling framework arising from either the MJP or CLE is fairly flexible and consequently, has been used ubiquitously in areas such as epidemiology \citep{oneill1999,lin2013b,mckinley2014}, population ecology \citep{BWK08,sun15} and systems biology \citep{Owen:2015,georgoulas17,GoliBrad19}.  

In order for the modelling framework to be of practical use, plausible parameter values must be obtained given data at discrete times, that may be incomplete (in the sense of information on a subset of species in the reaction network) and subject to error. This setting, when combined with either the MJP or CLE modelling framework precludes straightforward likelihood-based inference owing to the intractability of the observed data likelihood. Various sampling based solutions to this problem have been proposed including the use of data augmentation \citep[e.g.][]{BWK08} and approximate Bayesian computation \citep[ABC, e.g.][]{Owen15,Wu14}. 
Since it is straightforward in principle to unbiasedly estimate the observed data likelihood with a particle filter, particle Markov chain Monte Carlo \citep[particle MCMC,][]{andrieu09} provides another  sampling-based solution to the intractable likelihood problem, and has been widely adopted \citep[e.g.][]{mckinley2014,GoliWilk15,koblents2015}. Several modifications to the simplest particle MCMC scheme have been found to increase computational efficiency e.g. use of guided proposals inside the bootstrap particle filter \citep{Goli19,GoliBrad19}, correlated particles \citep{dahlin2015,deligiannidis2018} and use of a computationally cheap surrogate to prune out parameter proposals that are likely to be rejected giving a delayed acceptance scheme \citep{Goli15,quiroz18,Banterle19}.      

Herein, it is recognised that the above acceleration techniques typically leverage the tractability of a common surrogate model, and a novel combination and extension of these ideas are proposed to give a unified inferential framework. Specifically, the surrogate is constructed as a linear noise approximation \citep[see e.g.][]{kampen2001,Komorowski09,fearnhead14} which can be directly derived from either the MJP or CLE to give a linear Gaussian approximation of the transition probability governing the SKM. The surrogate is used in three ways: firstly, in the design of a Metropolis adjusted Langevin algorithm \citep[MALA, see e.g.][]{RobeTweed96}; secondly, to construct an appropriate bridge for use in the bootstrap particle filter; thirdly, in the first stage of a delayed-acceptance step, which is proven to be compatible with the use of correlated particles. The LNA requires the (numerical) solution of an ordinary differential equation (ODE) system whose number of components grows quadratically with the number of species. An efficient implementation that avoids unnecessary repeat runs of the ODE system is therefore considered. Crucially, the proposed framework exactly targets the posterior under either the CLE or MJP rather than under the surrogate model. The performance of the resulting framework is compared with existing approaches in three examples involving either the MJP or CLE as the inferential model.

The remainder of this paper is organised as follows. The modelling framework (including the MJP, CLE and LNA) is described in Section~\ref{sec:skm}. Section~\ref{sec:bayesInf} gives the inference task, whilst Section~\ref{sec:acc} provides details of the proposed acceleration techniques. Applications are considered in Section~\ref{sec:apps} and conclusions drawn in Section~\ref{sec:disc}.

\section{Stochastic kinetic models}
\label{sec:skm}
Consider a chemical reaction network (CRN) with $s$ species (also known as reactants) and $r$ reactions, defined by the tuple 
$\mathcal{N}=(A,B,\mathcal{X},c)$ with components as follows.
We have that $\mathcal{X}=(\mathcal{X}_1,\mathcal{X}_2,\ldots,\mathcal{X}_s)'$ is an $s$-vector of species, and
$A$ and $B$ are $r\times s$ matrices whose respective elements $\{a_{ij}\}$ and 
$\{b_{ij}\}$ are non negative integers known as stoichiometric coefficients. 
Additionally, $c=(c_1,\ldots,c_r)'$ is an $r$-vector of rate constants associated with an ordered reaction list given by 
\[
 A\mathcal{X} \xrightarrow{\phantom{fl}c\phantom{fl}} B\mathcal{X}.
\]
It should then be clear that $a_{ij}$ denotes the number of molecules of $\mathcal{X}_j$ consumed by reaction $\mathcal{R}_i$ 
and $b_{ij}$ denotes the number of molecules of $\mathcal{X}_j$ produced by reaction $\mathcal{R}_i$; the elements of $A\mathcal{X}$ are typically termed complexes. The effect of all reactions on all 
species can be encoded succinctly in the $s\times r$ stoichiometry matrix $S=(B-A)'$ so that, 
for example, the $i$th column of $S$, denoted $S_{.i}$, gives the change in the number of each 
species upon the occurrence of reaction $i$.

Let $X_{j,t}$ denote the (discrete) number of species $\mathcal{X}_j$ at time
$t$, and let $X_t$ be the $s$-vector $X_t = (X_{1,t},X_{2,t},\ldots,\linebreak[0] X_{s,t})'$. 
We model the process $\{X_t,t\geq 0\}$ via a Markov jump process (MJP), so that the state of the system at time $t$ is
\begin{equation}\label{MJP1}
X_t=x_0+\sum_{i}S_{.i}R_{i,t},
\end{equation}
where $x_0$ is the initial system state and each $R_{i,t}$, $i=1,\ldots,r$, is a counting process with intensity $h_i(x_t)$, known in 
this setting as the reaction hazard, which depends explicitly on the current state of the system $x_t$. 
Following \cite{kurtz1972} (see also \cite{Wilkinson06})
\begin{equation}\label{MJP2}
R_{i,t}=Y_i\left(\int_{0}^{t}h_i(x_{t'})dt' \right),
\end{equation}
where the $Y_i$, $i=1,\ldots,r$ are independent, unit rate Poisson processes. In what follows, 
the standard assumption of mass-action kinetics is assumed, so that 
\[
h_i(x_t) = c_i\prod_{j=1}^s \binom{x_{j,t}}{a_{ij}},
\] 
with explicit dependence on the rate constant $c_i>0$ omitted for notational convenience.

Given a value of the initial system state $x_0$ and rate constants $c$, 
exact realisations of the MJP can be generated via \emph{Gillespie's direct method} \citep{Gillespie77}. 
In brief, if the current state of the system is $x_t$, the time to the next reaction is $\tau\sim Exp(\sum_{i=1}^r h_i)$ 
and will be reaction $i$ with probability proportional to $h_i$, where $h_i:=h_i(x_t)$. Although the forward 
simulation is straightforward, the reverse problem is not. That is, the problem of inferring the rate 
constants $c$ given observations on $X_t$ at discrete times. The main barrier to inference in this 
setting arises from the intractability of the transition probability $p(x_t|x_0)$, which can be 
shown \citep{kampen2001} to satisfy the chemical master equation (CME):
\[
\frac{d}{dt} p(x_t|x_0) = \sum_{i=1}^r\left[h_i(x_t-S_{.i})p(x_t-S_{.i}|x_0)-h_i(x_t)p(x_t|x_0)\right].
\] 
Unfortunately, the CME can rarely be solved in practice, with the exactly solvable cases described 
in \cite{mcquarrie67}. Consequently, given data $\mathcal{D}=(x_{t_0},\ldots,x_{t_n})$, analytic 
evaluation of the \emph{observed data likelihood} $p(\mathcal{D}|c)$ is typically not possible. On the other hand, 
\emph{the complete data likelihood} \citep[see e.g.][]{Wilkinson06}, given by $p(x|c)$ where $x=\{x_t, t_0\leq t \leq t_n\}$, can be evaluated as 
\[
p(x|c)=p(x_{t_0})\left\{\prod_{i=1}^{n_{r}}h_{\nu_{i}}\left(x_{s_{i-1}}\right)\right\}
\exp\left\{-\int_{t_0}^{t_n}\sum_{i=1}^r h_{i}\left(x_t\right)dt\right\}.
\] 
Here, $n_{r}$ denotes the total number of reaction events; reaction times 
(assumed to be in increasing order) and types are denoted by 
$(s_{i},\nu_{i})$, $i=1,\ldots ,n_{r}$, $\nu_{i}\in \{1,\ldots ,r\}$ and we take $s_{0}=t_0$. Note that the exponent requires integration of the combined hazard function, which can be calculated analytically by recognising that the combined hazard function is piecewise constant in $x_t$. Although a complete data scenario is likely to be practically infeasible, 
the tractability of the complete data likelihood motivates simulation based approaches to inference 
based on data augmentation, whereby a sampler is constructed to target the joint posterior 
of $c$ and the latent jump process between observation instants, or uncertainty for the latent process, 
is integrated over via Monte Carlo. These techniques can be computationally prohibitive. Therefore, 
an approximation to the MJP is also considered, for which the computational cost can be controlled via time discretisation. 

\subsection{Discretised chemical Langevin equation} 
Consider an infinitesimal time interval, $(t,t+dt]$, over which the
reaction hazards will remain constant almost surely. Consequently, via 
(\ref{MJP2}), the counting process over this interval for the $i$th 
reaction, denoted by $dR_{i,t}$, is Poisson distributed 
with rate $h_i dt$. Stacking these quantities in $dR_t$ and noting 
that from (\ref{MJP1}) $dX_t=S dR_t$, it should be clear that
\[
\operatorname{E}(dX_t)=S\,h(x_t)dt,\qquad \operatorname{Var}(dX_t)= S\operatorname{diag}\{h(x_t)\}S'dt,
\]
where $h(x_t)=(h_1(x_t),\ldots,h_r(x_t))'$. Hence, an It\^o Stochastic differential equation (SDE) can be constructed, 
with an infinitesimal mean and variance that match those of the MJP. This is given by
\begin{equation}\label{cle}
dX_t = S\,h(x_t)dt + \sqrt{S\operatorname{diag}\{h(x_t)\}S'}\,dW_t,
\end{equation}
where $W_t$ is as $s$-vector of standard Brownian motion and $\sqrt{S\operatorname{diag}\{h(x_t)\}S'}$ 
is an $s\times s$ matrix $B$ such that $BB'=S\operatorname{diag}\{h(x_t)\}S'$. Equation \eqref{cle} 
is typically referred to as the chemical Langevin equation (CLE). The CLE can rarely be solved analytically, 
and it is common to work with a discretisation such as the Euler-Maruyama discretisation which gives
\[
X_{t+\Delta t} = x_t+ S\,h(x_t)\Delta t + \sqrt{S\operatorname{diag}\{h(x_t)\}S'\Delta t}\,Z,
\]
where $Z$ is a standard multivariate Gaussian random variable. However, the transition density
 $p_e(x_{t_{i+1}}|x_{t_i},c)$ under the Euler-Maruyama scheme is likely to be inaccurate unless 
$\Delta t = t_{i+1}-t_i$ is `small'. Hence, it is commonplace to introduce intermediate 
time points between observation instants allowing the discretisation to operate over a time step 
chosen by the practitioner. To this end, consider an equally spaced partition of $[t_i,t_{i+1}]$ 
as 
\[
t_i=\tau_{i,0}<\tau_{i,1}<\ldots < \tau_{i,m-1} < \tau_{i,m}=t_{i+1}, 
\]
with $\tau_{i,j+1}-\tau_{i,j}=\Delta\tau=1/m$ for $j=0,\ldots,m-1$. The value of $m$ is chosen to balance accuracy and computational efficiency, with a common practice being to perform short pilot runs with increasing values of $m$, until a threshold value is found whereby posterior output for any larger values of $m$ is approximately equal \citep[see e.g.][]{stramer11,GoliBrad19}. The transition density under this augmented Euler approach is
\[
p_{e}^{(m)}(x_{t_{i+1}}|x_{t_i},c)=\int p_{e}(x_{(t_i,t_{i+1}]}|x_{t_i},c)dx_{(t_i,t_{i+1})},
\]
where $x_{(t_i,t_{i+1})}=(x_{\tau_{i,1}},\ldots,x_{\tau_{i,m-1}})$ consists of grid points strictly between $t_i$ and $t_{i+1}$, and $x_{(t_i,t_{i+1}]}=(x_{\tau_{i,1}},\ldots,x_{\tau_{i,m}})$ includes $x_{t_{i+1}}$ itself. Finally, $p_{e}(x_{(t_i,t_{i+1}]}|x_{t_i},c)$ is a product of one step Euler transition densities over the intermediate time points. 
Then, given data $\mathcal{D}=(x_{t_0},\ldots,x_{t_n})$, the observed data likelihood is 
\[
p_{e}^{(m)}(\mathcal{D}|c)=p(x_{t_0})\prod_{i=0}^{n-1} p_{e}^{(m)}(x_{t_{i+1}}|x_{t_i},c),
\] 
which is typically intractable, owing to the intractability of the constituent terms. On the other hand, as with the MJP, the complete data 
likelihood $p_{e}(x|c)$, where $x=(x_{[t_{0},t_{1})},x_{[t_{1},t_{2})},\ldots, x_{[t_{n-1},t_n]})$, can be easily evaluated. 
Nevertheless, inference schemes based on $p_{e}(x|c)$ can be computationally prohibitive depending on the choice of $m$. This motivates a further approximation to the CLE (and therefore MJP) for which the observed data likelihood is tractable. 

\subsection{A tractable surrogate} \label{sec:lna}
The CLE can be appropriately linearised to give a tractable 
Gaussian process approximation known as the linear 
noise approximation \citep[LNA, see e.g.][]{kurtz1972,Komorowski09,stathopoulos13,fearnhead14}. 

Let $\{\eta_t, t\geq t_0\}$ be the deterministic process satisfying the ODE
\begin{equation}\label{lna1}
\frac{d\eta_t}{dt} = S h(\eta_t),\qquad \eta_{t_0}=\eta_{0},
\end{equation} 
and consider a residual stochastic process 
$\{R_t=X_t-\eta_t, t\geq t_0\}$ satisfying
\begin{equation}\label{eqn:resid_sde}
dR_t=\{Sh(x_t)-Sh(\eta_t)\}\,dt+\sqrt{S\operatorname{diag}\{h(x_t)\}S'}\,dW_t.
\end{equation}
This SDE can be approximated to give a tractable solution $\hat{R}_t$ and in turn 
$\hat{X}_t=\eta_t+\hat{R}_t$. This is obtained by Taylor expanding $S h(x_t)$ 
and $S\operatorname{diag}\{h(x_t)\}S'$ about $\eta_t$. Retaining the first two terms in the 
expansion of the former and the first term in the expansion of the latter gives
\begin{equation}\label{eqn:approx_resid_sde}
d\hat{R}_t=F_t\hat{R}_t\,dt+ \sqrt{S\operatorname{diag}\{h(\eta_t)\}S'}\,dW_t,
\end{equation}
where $F_t$ is the Jacobian matrix with ($i$,$j$)th element given by the partial derivative of the $i$th component of $S h(\eta_t)$ with respect to the $j$th component of $\eta_t$. 
 
Given an initial condition $\hat{R}_{t_0}\sim N(\hat{r}_0,\hat{V}_0)$, we obtain 
$\hat{R}_t$ as a Gaussian random variable. In particular, for the fixed value 
$\hat{r}_0=x_{t_0}-\eta_{t_0}$, it can be shown that the approximating distribution of $X_t$ is \[
X_t |X_{t_0}=x_{t_0} \sim N(\eta_t + G_t \hat{r}_{0}, G_t \psi_t G_t'),
\]
where $\eta$ satisfies (\ref{lna1}), the fundamental matrix $G_t$ satisfies
\begin{equation} \label{lna2}
\frac{dG_t}{dt} = F_t G_t, \quad G_{t_0} = I_s,
\end{equation}
and $\psi_t$ satisfies
\begin{equation} \label{lna3}
\frac{d\psi_t}{dt} = G_t^{-1}S\operatorname{diag}\{h(\eta_t)\}S'\left(G_t^{-1}\right)', \quad \psi_{t_0} = 0_s.
\end{equation}
Note that $I_s$ denotes the $s\times s$ identity matrix and $0_s$ the $s\times s$ zero matrix.

An equivalent representation of the LNA can be achieved by writing
\begin{equation} \label{Rsol2}
(\hat{R}_t|\hat{R}_{t_0} = \hat{r}_0) \sim N(m_t, V_t),
\end{equation}
where
\begin{equation} \label{lna2.5}
\frac{dm_t}{dt} = F_t m_t, \quad m_{t_0} = \hat{r}_0,
\end{equation}
and the ODE for $V_t = G_t \psi_t G_t'$ can be found using the product rule to give
\begin{equation} \label{lna3.5}
\frac{dV_t}{dt} = V_tF_t' + S\operatorname{diag}\{h(\eta_t)\}S' + F_t V_t, \quad V_{t_0} = V_0.
\end{equation}
The approximating distribution of $X_t$ for this alternative representation is
\[
(X_t|X_{t_0}=x_{t_0}) \sim N(\eta_t + m_t, V_t).
\]
Note that for the initial conditions $\eta_{t_0}=x_{t_0}$ and $m_{t_0}=0_s$, the ODE satisfied by $m_t$ need not be solved since $m_t=0_s$ for all $t\geq t_0$. A theoretical treatment of the LNA, and in particular, the conditions under which the LNA can be regarded as 
an adequate approximation to the CLE can be found in \cite{Wallace12}. The accuracy of the LNA is further discussed in an inferential setting in Section~\ref{sec:acc}.

\subsection{Example: Epidemic model}\label{sec:exEpi}
A susceptible--Infected--Removed \cite[SIR, e.g.][]{keeling07} compartment model has two species 
(susceptibles $\mathcal{X}_{1}$ and infectives $\mathcal{X}_{2}$) and two reaction channels (infection of 
a susceptible and removal of an infective):
\begin{align*}
\begin{array}{rrrl}
\mathcal{R}_1: & \quad \mathcal{X}_{1} + \mathcal{X}_{2} & \xrightarrow{\phantom{a}c_1\phantom{a}} & 2\mathcal{X}_{2}\\
\mathcal{R}_2: & \quad \mathcal{X}_{2} & \xrightarrow{\phantom{a}c_2\phantom{a}} & \emptyset.
\end{array}
\end{align*}
Let $X_{t}=(X_{1,t},X_{2,t})'$ denote the system state at time $t$. The stoichiometry matrix associated with the reaction system is given by
\[
S = \left(\begin{array}{rrr} 
 -1 & 0 \\
 1 & -1 
\end{array}\right)
\]
and the associated hazard function obtained under the assumption of mass-action kinetics is is 
\[
h(x_{t},c) = (c_1 x_{1,t}x_{2,t},c_2 x_{2,t})'.
\] 
The CLE for this model is given by
\[
d\begin{pmatrix}X_{1,t} \\X_{2,t} \end{pmatrix}=
\begin{pmatrix}-c_1 x_{1,t}x_{2,t} \\c_1 x_{1,t}x_{2,t}-c_2 x_{2,t} \end{pmatrix}\,dt+
\begin{pmatrix}c_1 x_{1,t}x_{2,t}   & -c_1 x_{1,t}x_{2,t} \\ -c_1 x_{1,t}x_{2,t} &c_1 x_{1,t}x_{2,t}+c_2 x_{2,t} 
\end{pmatrix}^{\frac{1}{2}}\,d\begin{pmatrix}W_{1,t} \\W_{2,t} \end{pmatrix}
\]
where $W_{1,t}$ and $W_{2,t}$ are independent standard Brownian motion processes. The LNA for this model is specified by the coupled ODE system given by (\ref{lna1}), (\ref{lna2}) and (\ref{lna3.5}), which requires the Jacobian matrix given by
\begin{align*}
F_t &= 
\begin{pmatrix}
 - c_1 \eta_{2,t} & -c_1 \eta_{1,t} \\
c_1 \eta_{2,t} & c_1 \eta_{1,t} - c_2
\end{pmatrix}.
\end{align*} 

Given initial conditions $x_0$ and values of the rate constants $c$, simulation (at discrete times) of with the MJP, CLE or LNA representation of the above reaction system is straightforward. The focus here is the \emph{inverse problem}, that is, given (assumed noisy) observations on (a subset of components of) $X_t$, plausible values of the rate constants $c$ are found. In what follows, the learning objective is described and the the proposed approach is outlined. 

\section{Bayesian inference}
\label{sec:bayesInf}

Suppose that the stochastic kinetic model (MJP or discretised CLE) is not observed directly, 
but observations (on a regular grid) $y_{t_i}, i=0,1,\ldots n$ are available 
and assumed conditionally independent (given the latent process), with conditional probability 
distribution obtained via the observation equation,
\begin{equation}\label{obs_eq}
Y_{t_i}=P'X_{t_i}+\varepsilon_{t_i},\qquad \varepsilon_{t_i}\sim \textrm{N}\left(0,\Sigma\right),\qquad i=0,1,\ldots, n.
\end{equation} 
Here, $Y_{t}$ is a length-$p$ vector, $P$ is a constant matrix of dimension 
$s\times p$ and $\varepsilon_{t}$ is a length-$p$ 
Gaussian random vector. The density linking the observed and latent process is denoted by 
$p(y_{t_i}|x_{t_i})$. For simplicity it is assumed that $\Sigma$ is known.

Let $\mathcal{D}=(y_{t_0},\ldots,y_{t_n})$ and suppose that $\pi(c)$ is the prior density ascribed to $c$. Throughout this article, it is  assumed that interest lies primarily in inference for the rate constants $c$. Therefore, the marginal parameter posterior density 
is constructed as 
\begin{align}\label{post}
\pi(c|\mathcal{D}) &\propto \pi(c)\int p(x|c) p(\mathcal{D}|x) dx \nonumber \\
 & \propto \pi(c)p(\mathcal{D}|c),
\end{align}
where $p(\mathcal{D}|x)=\prod_{i=0}^{n}p(y_{t_i}|x_{t_i})$ and 
$p(x|c)$ is the complete data likelihood under either the MJP or the discretised CLE 
(and no distinction between the two is subsequently made). As noted earlier, the intractability of the observed data likelihood $p(\mathcal{D}|c)$ complicates the inference task. In what follows a particle MCMC approach for generating draws from (\ref{post}) is described, and the resulting algorithm is used as a starting point for further acceleration techniques.

\subsection{Particle MCMC}\label{sec:pmcmc}

Given a non-negative estimator of $p(\mathcal{D}|c)$ that is unbiased up to a multiplicative constant (independent of the rate constants), the particle marginal Metropolis-Hastings (PMMH) scheme of \cite{andrieu09} targets a joint density for which the desired posterior $\pi(c|\mathcal{D})$ is a marginal.

Denote the estimator of the observed data likelihood by $\hat{p}_U(\mathcal{D}|c)$, where $U \sim g(u)$ is the set of random variables used to generate the estimator. The PMMH scheme targets the joint density
\begin{equation}\label{pspost}
\hat{\pi}(c,u|\mathcal{D}) \propto \pi(c)\hat{p}_u(\mathcal{D}|c)g(u),
\end{equation}
for which it is easily shown that $\pi(c|\mathcal{D})$ is obtained by integrating over the auxiliary variables $u$: 
\begin{align*}
    \int \hat{\pi}(c,u|\mathcal{D})du &= 
    \int \pi(c)\hat{p}_u(\mathcal{D}|c)g(u) du \\
    &= E_U[\pi(c)\hat{p}_u(\mathcal{D}|c)] \\
    & \propto \pi(c)p(\mathcal{D}|c).
\end{align*}

It remains to generate realisations of an unbiased estimator $\hat{p}_U(\mathcal{D}|c)$. The observed data likelihood can be factorised as
\begin{equation}\label{eqn:obslike}
p(\mathcal{D}|c) = p(y_{t_0}|c) \prod_{i=0}^n p(y_{t_{i+1}}| y_{t_0:t_i}, c),
\end{equation}
where $y_{t_0:t_i} = (y_{t_0},\ldots, y_{t_i})$. This immediately suggests a sequential approach for constructing  $\hat{p}_U(\mathcal{D}|c)$. The bootstrap particle filter \citep[see e.g.][]{gordon93} consists of a sequence of weighted resampling steps, whereby $N$ state particles are propagated forward, appropriately weighted using the complete data likelihood and observation density, and resampled with replacement \citep[e.g. systematically as in][]{deligiannidis2018} to prune out particle paths with low weight. The reader is referred to \cite{pitt12}; see also \cite{delmoral04} for a detailed explanation of the bootstrap particle filter as well as a proof of the requisite unbiasedness property; namely that the product (over time) of the average unnormalised particle weights gives an unbiased estimator of $p(\mathcal{D}|c)$.   

Step $i$ of the particle filter is given in Algorithm \ref{auxPF}. Note that the terms of the complete data likelihood $p(x_{(t_i,t_{i+1}]}^{(k)}|x_{t_i}^{(k)},c)$ will differ depending on whether the CLE or the MJP is used as the inferential model. A key ingredient of the algorithm is an appropriate construct $q(\cdot|x_{t_i},y_{t_{i+1}},c)$ for generating particle paths between observation instants $t_i$ and $t_{i+1}$.  Discussion of this construct is deferred to Section~\ref{sec:reduct}. 

\begin{algorithm}[t]
\caption{Step $i$ of the Particle Filter}\label{auxPF}
\textbf{Input}: Rate parameters $c$, next observation $y_{t_{i+1}}$, $N$ particles $\{x_{t_i}^{(k)}\}_{k=1}^N$.

\begin{enumerate}
\item Propagate forward to time step $t_{i+1}$ using 
\[
x_{(t_i,t_{i+1}]}^{(k)} \sim q(\cdot|x_{t_i}^{(k)}, y_{t_{i+1}}, c),
\]
for $k=1,\ldots,N$, using an appropriate proposal mechanism $q(\cdot|\cdot)$ (see Section \ref{sec:bridge} for details).
\item Compute the weights. For $k=1,\ldots,N$
\[
\tilde{w}_{t_{i+1}}^{(k)}=\frac{p(y_{t_{i+1}}|x_{t_{i+1}}^{(k)},c)p(x_{(t_i,t_{i+1}]}^{(k)}|x_{t_i}^{(k)},c)}
{q(x_{(t_i,t_{i+1}]}^{(k)}|x_{t_i}^{(k)},y_{t_{i+1}},c)}, \qquad w_{t_{i+1}}^{(k)}=\frac{\tilde{w}_{t_{i+1}}^{(k)}}{\sum_{j=1}^{N}\tilde{w}_{t_{i+1}}^{(j)}}.
\]
\item Resample $N$ particles using a systematic resampling step with weights $\{w_{t_{i+1}}^{(k)}\}_{k=1}^N$.
\end{enumerate}

\textbf{Output}: $N$ particles $\{x_{t_{i+1}}^{(k)}\}_{k=1}^N$ to be used in step $t_{i+1}$, an estimate for the current marginal likelihood term $\hat{p}_{u}(y_{t_{i+1}}|y_{t_0:t_i},c)=\frac{1}{N}\sum_{k=1}^{N}\tilde{w}_{t_{i+1}}^{(k)}$
\end{algorithm}

The PMMH scheme uses a proposal kernel $q(c^*|c)g(u^*)$, which, for the target in (\ref{pspost}) leads to an acceptance probability of the form
\[
\alpha\left(c^*|c\right) = \textrm{min} \left\{ 1, \frac{\pi(c^*) \hat{p}_{u^*}(\mathcal{D}|c^*)}{\pi(c) \hat{p}_{u}(\mathcal{D}|c)} \times \frac{q(c|c^*)}{q(c^*|c)} \right\}.
\]
Thus, each iteration of PMMH requires running a bootstrap particle filter with $N$ particles to obtain $\hat{p}_{u^*}(\mathcal{D}|c^*)$.  This can be computationally costly, since the number of particles should be scaled in proportion to the number of data points to maintain a desired variance of the logarithm of the likelihood estimator \citep{berard14}. 

\section{Acceleration techniques}
\label{sec:acc}

In this section, a unified inference framework that simultaneously aims to avoid unnecessary calculations of $\hat{p}_{u^*}(\mathcal{D}|c^*)$, reduce the variance of the likelihood estimator for a given $N$ and use a parameter proposal mechanism informed by an approximation of the marginal posterior density, is described. To facilitate these techniques, the tractability of a surrogate model is leveraged; this is the linear noise approximation (LNA) described in Section~\ref{sec:lna}. In what follows, therefore, the necessary surrogate preliminaries are described as well as their use for accelerating the particle MCMC scheme given in the previous section. An overview of the resulting algorithm, with reference to the techniques discussed in this section, can be found in Appendix~\ref{app:general}. Finally, it is worth emphasising that the role of the surrogate is to improve statistical and/or computational efficiency relative to the most basic particle MCMC scheme; the resulting algorithm exactly targets the posterior under the MJP or CLE. 

\subsection{Surrogate posterior and gradient information}
\label{sec:surrogate}

Denote the posterior under the LNA by
\[
\pi_\LNA(c|\mathcal{D}) \propto \pi(c)p_\LNA(\mathcal{D}|c)
\]
where the observed data (surrogate) likelihood $p_\LNA(\mathcal{D}|c)$ can be factorised as 
\begin{equation}\label{eqn:obslike2}
p_\LNA(\mathcal{D}|c) = p_\LNA(y_{t_0}|c) \prod_{i=0}^n p_\LNA(y_{t_{i+1}}| y_{t_0:t_i}, c).
\end{equation}
Constituent terms in (\ref{eqn:obslike2}) are tractable, and can be computed recursively using a forward filter; step $i$ of this approach can be found in Algorithm~\ref{algLNAff}. In brief, Bayes Theorem is applied sequentially by combining the Gaussian prior distribution of $X_{t_{i+1}}|y_{t_{1}:t_{i}}$ with the linear Gaussian observation equation (\ref{obs_eq}) to obtain a Gaussian posterior distribution of $X_{t_{i+1}}|y_{t_{1}:t_{i+1}}$ which is used to construct the prior at the next observation time. To alleviate potential inconsistencies between the LNA and CLE mean, which can be increasingly problematic over long inter-observation intervals, the algorithm of \cite{fearnhead14}
\citep[see also][for an alternative approach]{Minas17} is used; this approach re-initialises the LNA mean and variance at the mean and variance of the filtering distribution at the start of each inter-observation window. 

It is also necessary to compute the log gradient $\nabla \log \pi_\LNA(c|\mathcal{D})$ (see Section~\ref{sec:malaProp}). This requires $\nabla \log p_\LNA(\mathcal{D}|c)$ which we obtain by differentiating the logarithm of the terms in (\ref{eqn:obslike2}). Explicitly, from the observation model (\ref{obs_eq}),
\[
\nabla \log p_{LNA}(y_{t_{i+1}}|y_{t_0:t_i},c) = \nabla \log N(y_{t_{i+1}};P'\eta_{t_{i+1}},P'V_{t_{i+1}}P + \Sigma).
\]
For ease of notation, set $\mu(c,t) = P'\eta_{t+1}$ and $\Psi(c,t) = P'V_{t+1}P + \Sigma$, where $\eta_{t+1}$ and $V_{t+1}$ are both implicitly dependent on the rate parameters $c$. Then, 
\begin{equation} \label{dN/dt}
\frac{\partial \log N(y; \mu(c,t), \Psi(c,t))}{\partial c_i} = \frac{1}{2} \textrm{Tr} \left\{ (\gamma \gamma^T - \Psi^{-1}(c,t)) \frac{\partial \Psi(c,t)}{\partial c_i} \right\} + \gamma^T \frac{\partial \mu(c,t)}{\partial c_i}
\end{equation}
where $\gamma = \Psi^{-1}(c,t) \{y - \mu(c,t)\}$. Evaluating (\ref{dN/dt}) requires the partial derivatives $\partial \mu(c,t)/ \partial c_i$ and $\partial \Psi(c,t)/ \partial c_i$ which can be viewed as the first order sensitivities of the ODE system governing $\mu(c,t)$ and $\Psi(c,t)$. Although these are not in general available analytically, expressions for $d \mu(c,t)/ d t$ and $d \Psi(c,t)/ d t$ can be used to find expressions for the time derivatives of the first order sensitivities by augmenting the system of ODEs giving the LNA solution. Let $\xi$ be the $n_s$-vector of all elements of $\mu(c,t)$ and all lower triangular elements of $\Psi(c,t)$, and note that $n_s=s+s(s+1)/2$. The first order sensitivity of the $j$th element of $\xi$ with respect to the $i$th rate constant $c_i$ is given by 
\[
S_j^{(i)} = \frac{\partial \xi_j}{\partial c_i}, \quad j=1,\ldots, n_s, \quad i=1,\ldots,r.
\]
Due to the symmetry of second derivatives, the time derivatives of these sensitivities can be written as
\begin{equation}
\frac{d}{dt} S_j^{(i)}
= \sum_{l=1}^{n_s}S_l^{(i)} \frac{\partial }{\partial \xi_l}\frac{d \xi_j}{dt} + \frac{\partial}{\partial c_i}\frac{d \xi_j}{dt}, \quad j=1,\ldots, n_s, \quad i=1,\ldots,r.
\label{eq:sens}
\end{equation}
For further insight into first-order sensitivity equations, see \cite{cald11}. Given an initial condition of $S_j^{(i)} = 0$ at time $t_0$, these time derivatives can then be integrated forward numerically along with the rest of the component ODEs giving the LNA solution. Conveniently, calculation of $\nabla \log p_\LNA(y_{t_{i+1}}|y_{t_0:t_i},c)$ can be performed as part of the forward filter; see Algorithm~\ref{algLNAff}.

Finally, note that the augmentation of the LNA ODE system does come with an additional computational cost. With the addition of the sensitivity ODEs, the augmented ODE system has $(r+1)\left( s + s(s+1)/2 \right)$ ODEs in total to be solved, which can be computationally prohibitive for reaction systems with many species and/or reactions. This computational cost can be alleviated by making a further approximation and basing the gradient information solely on the deterministic part of the LNA. This is equivalent to ignoring the dependence of $\Psi(c,t)$ on $c$. The partial derivative in (\ref{dN/dt}) becomes
\begin{equation}
\frac{\partial \log N(y; \mu(c,t), \Psi(c,t))}{\partial c_i} = \gamma^T \frac{\partial \mu(c,t)}{\partial c_i},
\label{dN/dt_simp}
\end{equation}
thereby reducing the number of ODE components to $(r+1)s + s(s+1)/2$.

\begin{algorithm}[t]
\caption{Step $i$ of the LNA Forward Filter}\label{algLNAff}
\textbf{Input}: $a_{t_i}$ and $B_{t_i}$, the initial conditions of (\ref{lna1}) and (\ref{lna3.5}); $p_\LNA(y_{t_0:t_i}|c)$ and $\nabla \log p_\LNA(y_{t_0:t_i}|c)$, the current marginal likelihood and log gradient thereof; $y_{t_{i+1}}$, the next observation.
\begin{enumerate}
\item Prior at $t_{i+1}$. Initialise the LNA with $\eta_{t_i}=a_{t_i}$, $m_{t_i}=0$ and $V_{t_i}=B_{t_{i}}$. Integrate (\ref{lna1}), (\ref{lna3.5}) and (\ref{eq:sens}) forward to $t_{i+1}$ to obtain $\eta_{t_{i+1}}$, $V_{t_{i+1}}$, $\partial \mu(c,t)/ \partial c$ and $\partial \Psi(c,t)/ \partial c$. Thus
\[
(X_{t_{i+1}}|y_{t_1:t_i}) \sim N(\eta_{t_{i+1}},V_{t_{i+1}}).
\]
\item One step forecast. Using the observation equation, we have that
\[
(Y_{t_{i+1}}|y_{t_1:t_i}) \sim N(P'\eta_{t_{i+1}},P'V_{t_{i+1}}P + \Sigma).
\]
Hence compute
\[
p_\LNA(y_{t_0:t_{i+1}}|c) = p_\LNA(y_{t_0:t_i}|c)p_\LNA(y_{t_{i+1}}|y_{t_0:t_i},c)
\]
and
\[
\nabla \log p_\LNA(y_{t_0:t_{i+1}}|c) = \nabla \log p_\LNA(y_{t_0:t_i}|c)+ \nabla \log p_\LNA(y_{t_{i+1}}|y_{t_0:t_i},c).
\]
\item Posterior at $t_{i+1}$. Combining the distributions of $X_{t_{i+1}}$ and $Y_{t_{i+1}}$ (given $y_{t_0:t_i}$) and then conditioning on $y_{t_{i+1}}$ gives $(X_{t_{i+1}}|y_{t_0:t_{i+1}}) \sim N(a_{t_{i+1}},B_{t_{i+1}})$ where
\begin{align*}
a_{t_{i+1}} &= \eta_{t_{i+1}} + V_{t_{i+1}}P(P'V_{t_{i+1}}P + \Sigma)^{-1}(y_{t_{i+1}}-P'\eta_{t_{i+1}}) \\
B_{t_{i+1}} &= V_{t_{i+1}} - V_{t_{i+1}}P(P'V_{t_{i+1}}P + \Sigma)^{-1}P'V_{t_{i+1}}.
\end{align*}
\end{enumerate}
\textbf{Output}: $p_\LNA(y_{t_0:t_{i+1}}|c)$, $\nabla \log p_\LNA(y_{t_0:t_{i+1}}|c)$, $a_{t_{i+1}}$ and $B_{t_{i+1}}$.
\end{algorithm}

\subsection{Delayed-acceptance}

Consider now the particle MCMC scheme of Section~\ref{sec:pmcmc} targeting the $\hat{\pi}(c,u|\mathcal{D})$ in (\ref{pspost}) for which  $\pi(c|\mathcal{D})$ is a marginal. Ideally, iterations that run the particle filter to compute $\hat{p}_{u^*}(\mathcal{D}|c^*)$ when $c^*$ is likely to be rejected should be avoided. This motivates the use of a screening step, whereby the particle filter is only run for proposals accepted under the surrogate posterior. This is known as \emph{delayed-acceptance} (DA); a brief exposition is provided here and the reader is referred to relevant work \citep{christen2005,Goli15,Banterle19} for further details.

For a given iteration with current state $(c,u)$, Stage One of the DA scheme proposes $c^* \sim q(\cdot|c)$, computes $p_\LNA(\mathcal{D}|c^*)$ and the screening acceptance probability 
\begin{equation} \label{stage1}
\alpha_1\left(c^*|c\right) = \textrm{min} \left\{ 1, \frac{\pi(c^*) p_\LNA(\mathcal{D}|c^*)}{\pi(c) p_\LNA(\mathcal{D}|c)} \times \frac{q(c|c^*)}{q(c^*|c)} \right\}.
\end{equation}
If this screening step is successful, Stage Two of the DA scheme is to propose $u^* \sim g(\cdot)$, construct the estimate $\hat{p}_{u^*}(\mathcal{D}|c^*)$ and the Stage Two acceptance probability
\begin{align} \label{stage2}
\alpha_{2|1}\left\{(c^*,u^*)|(c,u)\right\} 
&= \textrm{min} \left\{ 1, \frac{\hat{p}_{u^*}(\mathcal{D}|c^*)}{\hat{p}_{u}(\mathcal{D}|c)} \times \frac{ p_\LNA(\mathcal{D}|c)}{ p_\LNA(\mathcal{D}|c^*)} \right\}.
\end{align}
Thus the overall acceptance probability for the scheme is
\begin{equation} \label{overall_acc_prob}
\alpha\left\{(c^*,u^*)|(c,u)\right\} = \alpha_1\left(c^*|c\right) \alpha_{2|1}\left\{(c^*,u^*)|(c,u)\right\},
\end{equation}
and standard arguments show that the resulting DA scheme defines a Markov chain that is reversible with respect to the target in (\ref{pspost}). Extensions of the two-stage approach that are robust to potentially poor choices of the surrogate can be found in \cite{Banterle19}.

Three modifications of the standard DA scheme (henceforth DA-PMMH) are now considered, with the aim of improving overall efficiency. These  are achieved through a combination of a gradient-based Stage One proposal and by reducing the variance of the Stage Two acceptance probability for a given number of particles $N$, for which two methods are suggested, that may be implemented separately or together.

\subsubsection{Stage One proposal}\label{sec:malaProp}

A common choice of proposal mechanism in MCMC schemes is the random walk Metropolis (RWM) proposal, in which
\[
c^* = c + \lambda Z,\qquad Z\sim \textrm{N}(0,\Sigma_T)
\]
for some tuning matrix $\Sigma_T$. For example it is common to take $\Sigma_T=\widehat{\textrm{Var}}(c|\mathcal{D})$ estimated from a pilot run, with $\lambda$ tuned to meet a desired acceptance rate. Further comments on tuning are given in Section~\ref{sec:tuning}.

It is desirable to find a proposal that uses local information about the
posterior to sample from areas of higher posterior density. The Metropolis adjusted Langevin algorithm (MALA) \citep[][]{Roberts02} incorporates the gradient of the log posterior density $\nabla \log \pi(c|\mathcal{D})$ in the proposal mechanism. As with RWM, a preconditioning matrix \citep[see e.g.][]{Marn2020} is included to give a proposal of the form
\[
c^* = c + \frac{\lambda^2}{2}\Sigma_T \nabla \log \left(\pi(c|\mathcal{D})\right) + \lambda Z, \quad Z \sim N(0,\Sigma_T).
\]
Unfortunately $\nabla \log \pi(c|\mathcal{D})$ is typically intractable. Estimates of this gradient can be computed via modification of the particle filter \citep{poyiadjis11,nemeth16}. However, and as discussed in \cite{nemeth16}, the asymptotic properties of the resulting algorithm depends crucially on the accuracy of the estimate of $\nabla \log \pi(c|\mathcal{D})$ as the number of parameters increases. This approach is therefore eschewed in favour of replacing the idealised gradient with the analytically tractable gradient of the log posterior under the surrogate model $\nabla \log \pi_\LNA(c|\mathcal{D})$, calculated using either (\ref{dN/dt}) (henceforth full MALA) or (\ref{dN/dt_simp}) (henceforth simplified MALA or sMALA). This is obtained through recursive application of Algorithm~\ref{algLNAff}.     

\subsubsection{Stage Two variance reduction via correlated particles}\label{sec:reduct} 
In this section, the recently proposed correlated PMMH (CPMMH) method \citep{dahlin2015,deligiannidis2018} is adapted to the delayed acceptance setting (DA-CPMMH). The idea is to induce strong and positive correlation between successive values of the likelihood estimates, as used in the Stage Two acceptance probability. Use of correlation within PMMH schemes in a delayed acceptance framework has been studied before by \cite{quiroz18}, who utilised the block pseudo-marginal approach of \cite{tran2016} to induce correlation between the likelihood estimates. By contrast, we  correlate the auxiliary random variables $u$ used to construct the estimator $\hat{p}_{u}(\mathcal{D}|c)$, by replacing the proposal $u^*\sim g(u^*)$ with $u^*\sim K(u^*|u)$ where the kernel $K(\cdot|\cdot)$ satisfies the detailed balance equation
\begin{equation} \label{gK_db}
g(u)K(u^*|u) = g(u^*)K(u|u^*).
\end{equation}
Without loss of generality, take $g(u)=\textrm{N}\left(u;\,0\,,\,I_d\right)$ where $d$ denotes the number of components of $u$, and $K(u^*|u)$ to be the density associated with a Crank-Nicolson proposal. That is
\begin{equation}
K(u^*|u)=\textrm{N}(u^*;\,\rho u\,,\,\left(1-\rho^2\right)I_d),
\label{eq:crank}
\end{equation}
where $\rho$ is a tuning parameter between $0$ and $1$ that determines the correlation between $u^*$ and $u$. Setting $\rho=0$ gives the standard DA-PMMH scheme, as in this case $K(u^*|u) = g(u^*)$. However, in practice $\rho$ is generally taken to be close to $1$, so as to induce strong positive correlation between successive estimates from the particle filter. It is expected that the resulting reduction in the variance of the Stage Two acceptance probability results in far fewer particles required for DA-CPMMH than for DA-PMMH, significantly reducing the relative computational cost. The use of correlation here is likely to be of most benefit in low dimensional models, since it is likely that $N$ can be scaled at rate $n^{1/2}$ for univariate models and $n^{2/3}$ for bivariate models \citep{deligiannidis2018}, as opposed to at rate $n$ for the standard PMMH scheme \citep{berard14}. 

Appendix~\ref{valid_app} shows that a delayed acceptance scheme with proposal kernel $q(c^*|c)K(u^*|u)$ and acceptance probability given by (\ref{overall_acc_prob}) satisfies detailed balance with respect to the target density (\ref{pspost}). 

\subsubsection{Stage Two variance reduction via bridge constructs}\label{sec:bridge}

Algorithm \ref{auxPF} requires the use of a proposal mechanism that can generate paths between observations for the particles, conditional on the current state of the particle, the next observation and the rate constants. These paths are often referred to as bridges, and the mechanisms for generating them are known as bridge constructs. \cite{gordon93} originally proposed generating particles via forward simulation from the model from one time point to the next, without taking into consideration the observation at the end time point. This method is termed the myopic approach. However, as explored in \cite{delmoral2015} and \cite{Goli15} (see also \cite{GoliBrad19}), when the observation variance is small relative to the intrinsic stochasticity exhibited by the latent process, this implementation can lead to a highly variable estimator of the marginal likelihood. In this case, as the precision of an observation increases, its compatibility with most of the paths reduces, leading to low weights, and the efficiency of bootstrap particle filter-driven (C)PMMH scheme decreases  substantially. By conditioning on the next observation, bridge constructs play an important role in reducing the variance of $\hat{p}_U(\mathcal{D}|c)$ relative to this myopic approach.

Without loss of generality, consider a time interval $(0,T]$ for which we require a bridge construct with density $q(x_{(0,T]}|x_{0},y_{T},c)$. Consider first the MJP as the inferential model and suppose that we have simulated as far as time $t$. A suitable bridge construct can be found by noting the conditioned hazard (CH) associated with reaction $\mathcal{R}_i$ is
\[
h_{i}(x_t|y_T)=h_{i}(x_t)\frac{p(y_T|X_{t}=x')}{p(y_T|X_t=x_t)},
\]
where $x'=x_t+S_{.i}$. The transition density $p(y_T|x_t)$ will typically be intractable and we follow \cite{Goli19} by replacing it with the transition density under the surrogate model
\[
p_\LNA(y_T|X_t=x_t)=\textrm{N}(y_T; P'[\eta_{T|0}+G_{T|t}(x_t-\eta_{t|0})]\,,\, P'V_{T|t}P+\Sigma).
\]
Here, the notation $\eta_{t'|t}$, $G_{t'|t}$ and $V_{t'|t}$ is used to denote the solution of the ODE system in (\ref{lna1}), (\ref{lna2}) and (\ref{lna3.5}) at time $t'$, integrated over $(t,t']$ with initial conditions $z_t = x_t$, $G_{t}=I_s$ and $V_t=0_s$. A single integration of the ODE system over $[0,T]$ gives $\eta_{t|0}$, $G_{t|0}$ and $V_{t|0}$ for $t\in [0,T]$. Then, obtain $G_{T|t}$ and $V_{T|t}$ via the identities
\begin{equation}\label{eq:identities}
G_{T|t}=G_{T|0}G_{t|0}^{-1}, \qquad V_{T|t}=V_{T|0}-G_{T|t}V_{t|0}G_{T|t}',
\end{equation}
which are derived in \cite{Goli19}. Use of (\ref{eq:identities}) avoids reintegration of the ODE system at each jump event. By ignoring the explicit dependence of $h_{i}(x_t|y_T)$ on $t$, sampling the resulting bridge proposal $q(x_{(0,T]}|x_{0},y_{T},c)$ can be achieved by executing Gillespie's direct method with $h_{i}(x_t)$ replaced by $h_{i}(x_t|y_T)$. Evaluating $q(x_{(0,T]}|x_{0},y_{T},c)$ is straightforward via the complete data likelihood of $x_{(0,T]}$, again with $h_i(x_t)$ replaced by the conditioned hazard function.

Consider now the discretised CLE as the inferential model. In this case, $x_{(0,T]}$ denotes the process over equally spaced  intermediate times $\tau_{1},\ldots,\tau_{m}=T$ with time step $\Delta\tau$, given an initial value $x_0=x_{\tau_0}$. Herein, the residual bridge construct of \cite{whitaker2017} is adopted. In brief, we partition $X_t$ as $X_t=\zeta_t+R_t$ where
\begin{align*}
d\zeta_t &= f(\zeta_t) dt , \qquad \zeta_0=x_0, \\
dR_t &= \left\{Sh(X_t)-f(\zeta_t) \right\}dt +\sqrt{S\operatorname{diag}\{h(X_t)\}S'} dW_t , \qquad R_0=0,
\end{align*}
for some function $f(\cdot)$. Although the conditional distribution $R_{\tau_{k+1}}|r_{\tau_k},y_T$ will necessarily be intractable (even under discretisation), a tractable linear Gaussian approximation can constructed \citep{whitaker2017}. Full details are given in Appendix~\ref{app:rbridge}. Finally, two choices of the function $f(\cdot)$ are considered. The first is $f(\cdot)=Sh(\cdot)$ with $\zeta_t=\eta_t$ giving the simple residual bridge (RB). The second choice has $\zeta_t=\eta_t+\hat{\rho}_t$, where $\hat{\rho}_t = \operatorname{E}[\hat{R}_t|r_0,y_T]$ is a surrogate approximation of the conditional expected residual at time $t$. Using the LNA, $\hat{\rho}_t$ is computed as  
\[
\hat{\rho}_t = G_{t|0}{r}_0 + V_{t|0}(G_{t|0}')^{-1}G_{T|0} P(P'V_{T|0} P + \Sigma)^{-1}(y_T - P'\eta_{T|0} - P'G_{T|0}{r}_0).
\]
The derivation of this expectation is given in Appendix~\ref{app:rbridge}. The resulting bridge construct is referred to as the residual bridge with extra subtraction (RB$^-$).

\subsection{Computational considerations}
Use of the surrogate LNA model in a delayed-acceptance step, the MALA parameter proposal and to construct bridge proposals inside the particle filter each require the solution of an ODE system. However, there is some overlap in the ODE components that must be solved to perform each technique, and as such, if implemented correctly, further computational savings can be made when using several of these techniques at once. 

Computing the observed data likelihood under the LNA for use in a delayed-acceptance step requires the solution of (\ref{lna1}) and (\ref{lna3.5}), restarted at the posterior mean and variance given by the forward filter at each observation time. Computing the gradient information to use full MALA requires the solution of (\ref{lna1}) and (\ref{lna3.5}), as well as the first order sensitivities $\partial \mu(c,t)/ \partial c_i$ and $\partial \Psi(c,t)/ \partial c_i$ for $i=1,\ldots,r$; see (\ref{eq:sens}). The gradient information using simplified MALA does not require the solution of $\partial \Psi(c,t)/ \partial c_i$. The simple residual bridge, $\textrm{RB}$, requires only the solution of (\ref{lna1}). The residual bridge with additional subtraction, $\textrm{RB}^-$, and conditioned hazard, CH, require the solution of 
(\ref{lna1}), (\ref{lna2}) and (\ref{lna3.5}). 

Except for RB, all of these techniques require the solution of (\ref{lna1}) and (\ref{lna3.5}). Thus, it is desirable to solve these ODEs \emph{once per (C)PMMH iteration} and use the output in several different techniques. Running the forward filter to obtain the surrogate likelihood used in delayed acceptance also solves several of the ODE components used in determining the gradient of the log posterior for MALA. Care must be taken when implementing the bridge constructs, which, for an arbitrary observation interval $[t_i,t_{i+1}]$ and time $t\in(t_i,t_{i+1}]$, require the LNA variance $V_{t|t_i}$ initialised at $0_s$, whereas the forward filter restarts this variance at the filtering mean $B_i$ (see Algorithm~\ref{algLNAff}). This ``disconnect'' is alleviated via the second identity in (\ref{eq:identities}) which can be written as 
\[
V_{t|t_i}=V_{t}-G_{t}V_{t}G_{t}'  
\]
where $V_t$ and $G_t$ are obtained from the forward filter. The resulting bridge constructs in this setting are denoted by $\textrm{RB}_\textrm{iter}$, $\textrm{RB}_\textrm{iter}^-$ and $\textrm{CH}_\textrm{iter}$. The accuracy of the bridges over $[t_i,t_{i+1}]$ can be improved by re-integrating the ODE system given by (\ref{lna1}) and (\ref{lna3.5}) \emph{for each particle} $x_{t_i}^{(k)}$. That is, $\eta_{t_i}$ is set at $x_{t_i}^{(k)}$ and $V_{t_i}=0_s$. The resulting bridge constructs are denoted by $\textrm{RB}_\textrm{part}$, $\textrm{RB}_\textrm{part}^-$ and $\textrm{CH}_\textrm{part}$. Although use of the latter compared to the ``once per iteration'' approach is likely to result in an estimator of observed data likelihood with lower variance and in turn, better mixing of the (C)PMMH scheme, it comes with an additional computational cost. Given $s$ species and $N$ particles, ``once per particle'' bridges require the solution of an additional $sN$ ODE components. Table~\ref{tab:complexity} shows the relative computational complexity (in terms of the number of ODE components that must be solved) for different acceleration techniques. Note that $\textrm{CH}_\textrm{iter}$ and $\textrm{CH}_\textrm{part}$ have the same computational complexities as $\textrm{RB}_\textrm{iter}^-$ and $\textrm{RB}_\textrm{part}^-$.

\begin{table}
    \centering
    \begin{tabular}{@{}lllll@{}}
    \toprule
     & (C)PMMH & da(C)PMMH & Simplified MALA & Full MALA \\
     \midrule
    $\textrm{RB}_\textrm{iter}$ & $O(s)$ & $+O(s^2)$ & $+O(sr)$ & $+O(s^2r)$ \\
    $\textrm{RB}_\textrm{iter}^-$ & $O(s^2)$ & $-$ & $+O(sr)$ & $+O(s^2r)$ \\
    $\textrm{RB}_\textrm{part}$ & $O(sN)$ & $+O(s^2)$ & $+O(sr)$ & $+O(s^2r)$ \\
    $\textrm{RB}_\textrm{part}^-$ & $O(s^2N)$ & $+O(s^2)$ & $+O(sr)$ & $+O(s^2r)$
    \end{tabular}
    \caption{Order of complexity in terms of ODE components required to be solved for different bridge construct implementations, and the additional computational cost required to enact delayed acceptance, simplified or full MALA. Note that $N$, $s$ and $r$ denote the number of particles, species and parameters respectively.}
    \label{tab:complexity}
\end{table}

\subsection{Tuning}\label{sec:tuning}
Schemes employing CPMMH require specification of a correlation parameter $\rho$, and irrespective of the acceleration technique employed, all schemes require specification of several other tuning parameters. These include a number of particles $N$, a preconditioning matrix $\Sigma_T$ and scaling parameter $\lambda$, with the latter two tuning parameters used in the RWM or MALA proposal mechanism. As the rate constants $c$ must be strictly positive, their natural logarithms are used when generating proposed values, by applying RWM and MALA to $\log c$. Hence, all schemes take $\Sigma_T=\widehat{\textrm{Var}}(\log c|\mathcal{D})$ estimated from a short pilot run, and, using several short pilot runs, find a value of $\rho$ that gives an effective sample size (ESS) value for the auxiliary variable chain consistent with the minimum (over parameter chains) ESS value (mESS). If several permissible values of $\rho$ are found, the value that gives the pilot run with the largest mESS (over parameter chains) is chosen. 

Practical advice for choosing the number of particles $N$ for PMMH can be found in \cite{Doucet2015} and \cite{sherlock2015}; see also \cite{schmon2021} for parameter dimension guidelines. For CPMMH, the guidance in \cite{deligiannidis2018} is used by choosing $N$ so that the variance of the logarithm of the ratio $\hat{p}_{u^*}(\mathcal{D}|c) / \hat{p}_{u}(\mathcal{D}|c)$ is around 1 with $c$ set at some central posterior value. For RWM, the guidance in \cite{schmon2021} is used by setting $\lambda$ to give an empirical acceptance rate of around $20\%$, depending on the number of parameters to be inferred. When using MALA, the practical advice of \cite{nemeth16} is applied and by aiming for an acceptance rate of around $40\%-50\%$. Guidance on tuning delayed acceptance (RWM) schemes can be found in \cite{sherlock2021}. For DA-CPMMH schemes with either a RWM or MALA proposal, the  number of particles is chosen by following the procedure above, and then conditional on this choice, the scaling is tuned to optimise mESS. Finally, with this scaling, the number of particles is chosen to optimise overall efficiency (in terms of minimum effective sample size per second).    

\section{Applications}\label{sec:apps}
In what follows, all algorithms are coded in R and were run on a desktop computer with an Intel quad-core CPU. 
For all experiments, the performance of competing algorithms are compared using minimum (over each parameter chain) effective sample size per second (mESS/s), computed using the R coda package \citep{Plummer06} and wall clock computing time. The latter is based on main monitoring
runs of the MCMC scheme considered and it is noted that the CPU cost of tuning was small relative to the cost of the main run and comparable across competing schemes. Marginal posterior densities are estimated using the output of the best mixing scheme. When using the discretised chemical Langevin equation as the inferential model (second and third application), $\Delta\tau$ was fixed at 0.1, which gave a reasonable balance between accuracy and computational efficiency. Further details regarding the form of the model employed in each application can be found in Section~\ref{sec:exEpi} and Appendix~\ref{app:models}. Computer code to implement all methods can be downloaded from \url{https://github.com/tl1995/daCPMMH_MALA}.

\subsection{Epidemic model}

Consider the well studied Eyam plague data set \citep[see e.g.][]{Raggett82} consisting of 8 observations on susceptible and infective individuals during the outbreak of plague in the village of Eyam, England, taken over a four month period from June 18th 1666. These data are presented here in Table~\ref{tab:tabE}.

\begin{table*}[t]
  \centering
  \small
  \begin{tabular}{@{} lllllllll@{}}
\toprule
& \multicolumn{8}{@{}c}{Time (months)}\\
                 &0  &0.5  &1  &1.5  &2   &2.5  &3  &4 \\
\midrule    
    Susceptibles &254  &235  &201  &153  &121   &110  &97  &83 \\
    Infectives   &\phantom{00}7  &\phantom{0}14  &\phantom{0}22  &\phantom{0}29  &\phantom{0}20   &\phantom{00}8  &\phantom{0}8  &\phantom{0}0  \\ 
\bottomrule
  \end{tabular}
  \caption{Eyam plague data.}\label{tab:tabE}
\end{table*}

For this application, the Markov jump process representation of species dynamics was taken as the inferential model. The challenging scenario of exact observation of all model components (albeit at discrete times) was assumed, for which the particle filter in Algorithm~\ref{auxPF} assigns a non zero weight to the particle $x_{(t_i,t_{i+1}]}^{(k)}$ if and only if $x_{t_{i+1}}^{(k)}$ is equal to the observation $y_{t_{i+1}}$. That is, simulated trajectories must ``hit'' the observation or else receive zero weight. In this exact observation setting, no resampling is required and the particle filter coincides with a series of independent importance samplers (over each observation interval). Hence, the ODE solution required to implement the conditioned hazard approach of Section~\ref{sec:bridge} need not be re-initialised for each particle and therefore $\textrm{CH}_\textrm{iter}$ and $\textrm{CH}_\textrm{part}$ coincide.   

\begin{figure}[ht]
\centering
\includegraphics[width=15cm,height=7cm]{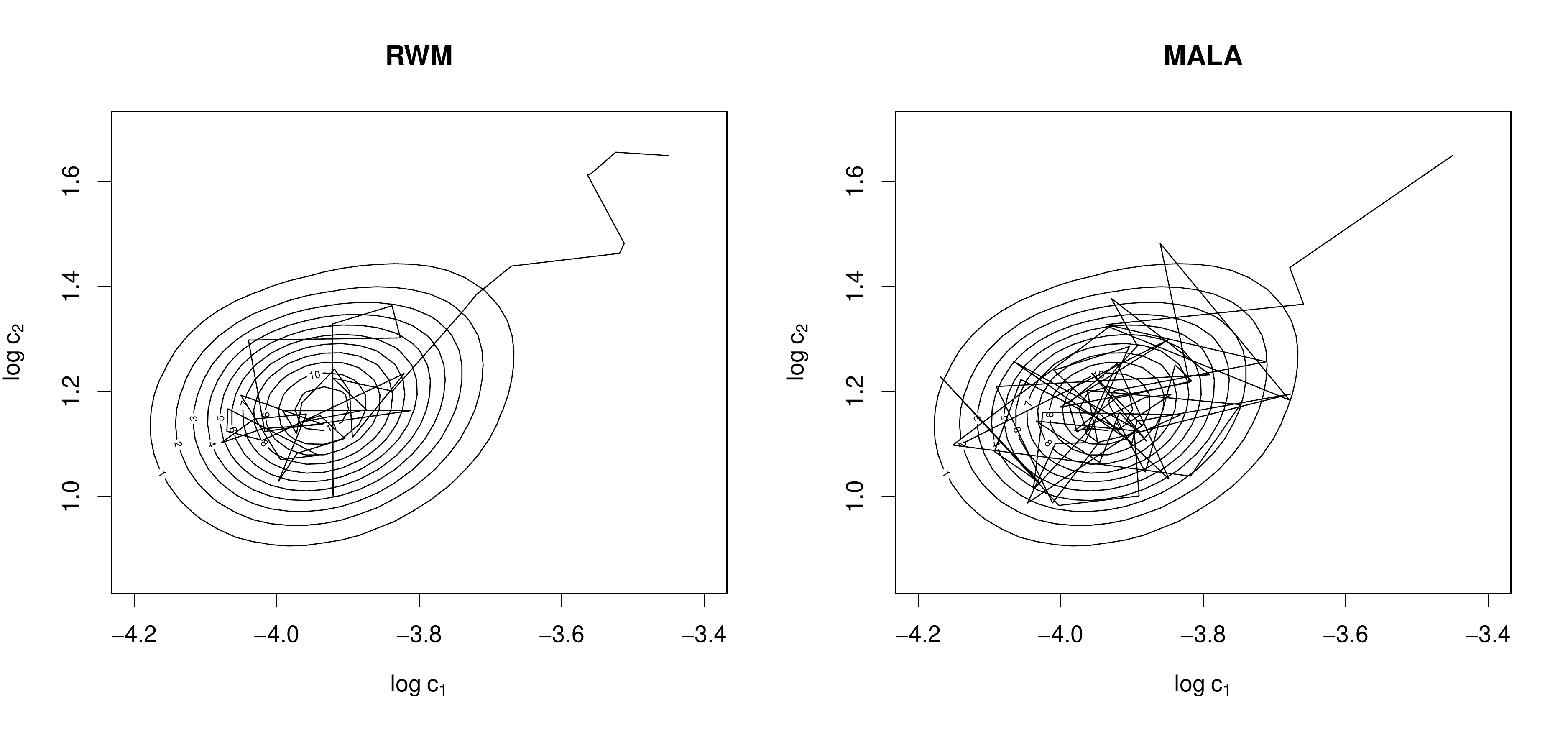}
\caption{Epidemic model. Joint posterior density and the first 100 iterations of CPMMH-RWM (left) and CPMMH-MALA (right).}
\label{fig:epi1}
\end{figure}

The independent prior specification of \cite{Ho2018} was used, with a $N(0,10^2)$ distribution assigned to the logarithm of each rate constant. The bridge-based CPMMH ($\rho=0.99$) was run with and without MALA, with and without delayed acceptance. For bench-marking, the standard PMMH (based on forward simulation, denoted ``Myopic'') and bridge-based PMMH were also run. The main monitoring runs consisted of $10^4$ iterations and this output is summarised in Table~\ref{tab:E2}. Use of the conditioned hazard and correlating reaction times / types between successive runs of the particle filter gives a modest improvement in overall efficiency (by a factor of 3) compared to the most basic PMMH scheme. For this particular target posterior (see Figure~\ref{fig:epi1}) MALA is clearly more effective than RWM and is more than twice as efficient (in terms of minimum ESS per second) compared to RWM. Combining CPMMH, delayed acceptance and MALA gives the best performing scheme. The increased performance due to delayed acceptance is unsurprising, given the accuracy of the surrogate (as evidenced by the Stage-Two acceptance probability) and its computational efficiency (with a relative cost of calculating the observed data likelihood under the surrogate versus an estimate from the particle filter scaling of around 1:100).   

\begin{figure}[ht]
\centering
\includegraphics[width=15cm,height=7cm]{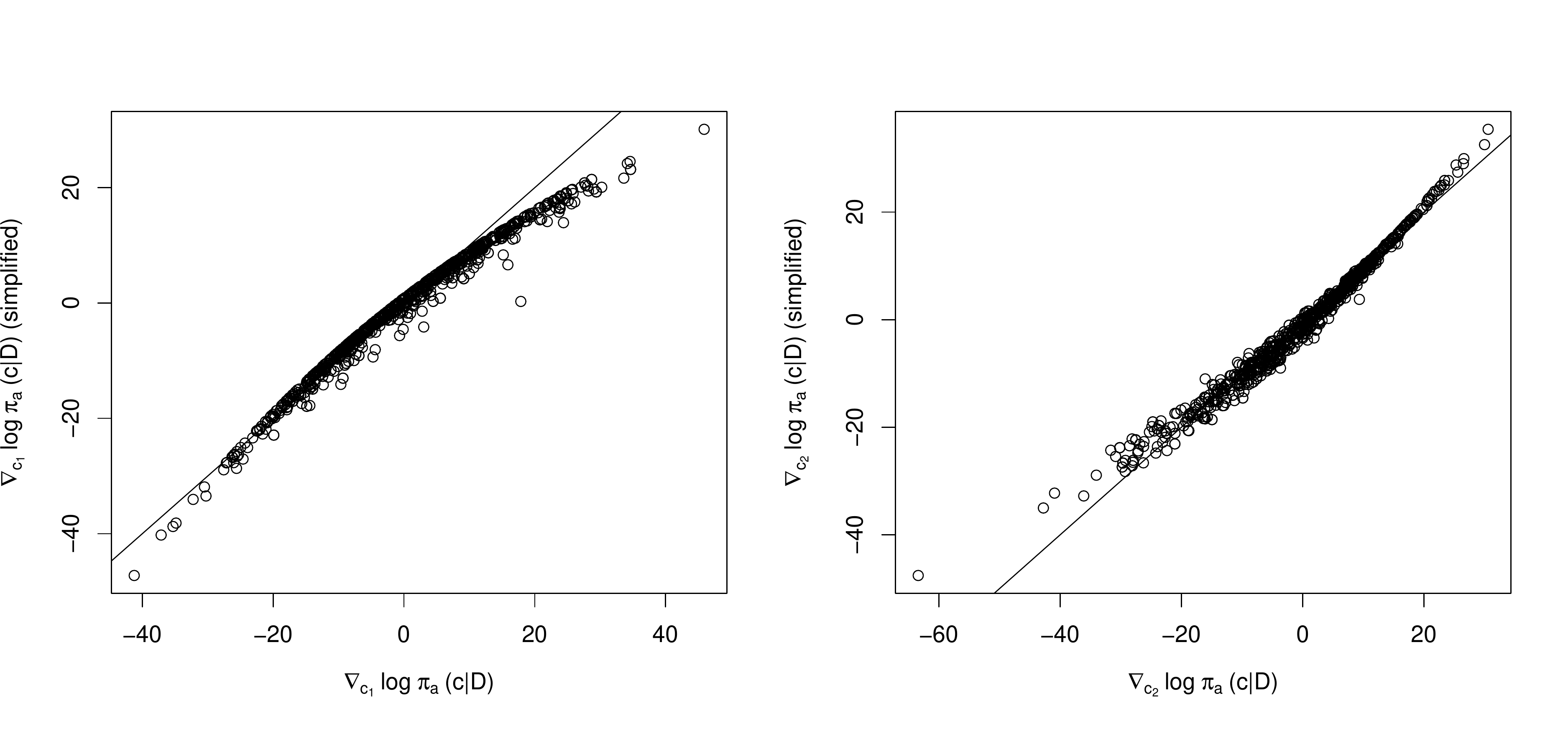}
\caption{Epidemic model. Full versus simplified gradient of the log posterior density with respect to $c_1$ (left) and $c_2$ (right) computed for $1000$ draws from the joint posterior over $c$.}
\label{fig:epi2}
\end{figure}

Finally, Figure~\ref{fig:epi2} and Table~\ref{tab:E2} suggest that although simplified MALA (sMALA, using equation (\ref{dN/dt_simp})) gives gradients of the log posterior that are generally comparable to full MALA (using equation (\ref{dN/dt})), the reduction in CPU time is negligible, and not sufficient to overcome the reduction in mixing efficiency. This is unsurprising given that CPU time is dominated by the particle filter, as noted above.

\begin{table}
\centering
\small
	\begin{tabular}{@{}lllllllll@{}}
         \toprule
Scheme & $N$ & $\alpha_{1}$ & $\alpha_{2|1}$ & $\alpha$ & CPU (s) & mESS & mESS/s & Rel.  \\
\midrule
PMMH / RWM (Myopic) & 5000 & -- & -- & 0.16 & 68177 & \phantom{0}863 & 0.013 & \phantom{0}1.0 \\
PMMH / RWM & \phantom{0}100 & -- & -- & 0.19 & 25752 & \phantom{0}644 & 0.025 & \phantom{0}2.0  \\
\midrule
CPMMH / RWM  & \phantom{00}75  & -- & -- & 0.25 & 15796 & \phantom{0}609 & 0.039 & \phantom{0}3.0 \\
CPMMH / MALA & \phantom{00}75 & -- & -- & 0.41 & 16040 & 1360 & 0.085 & \phantom{0}6.7 \\
CPMMH / sMALA & \phantom{00}75 & -- & -- & 0.42 & 15799 & \phantom{0}891 & 0.056 & \phantom{0}4.4 \\
\midrule
daCPMMH / RWM & \phantom{00}75 & 0.28 & 0.49 & 0.13 & \phantom{0}4746 & \phantom{0}340 & 0.071 & \phantom{0}5.6  \\
daCPMMH / MALA & \phantom{00}75 & 0.15  & 0.45 & 0.07 & \phantom{0}2840 & \phantom{0}386 & 0.136 & 10.7  \\
\bottomrule
\end{tabular}
      \caption{Epidemic model. Number of particles $N$, acceptance rates $\alpha_1$, $\alpha_{2|1}$ and $\alpha$, CPU time (in seconds), minimum ESS, minimum ESS per second, and relative (to the worst performing scheme) minimum ESS per second. All results are based on $10^4$ iterations of each scheme.}\label{tab:E2}
\end{table}

\subsection{Aphid model}

Aphids, also known as greenflies, are small, sap-sucking insects that feed on plants, often on the underside of leaves. Cotton aphids (\emph{Aphis gossypii}) are a species of aphid that are hosted on several plants, including cotton. When aphids initially infest a plant, they tend to reproduce far faster than they die. However, as well as damaging the plant directly, they also secrete honeydew over the plant leaf, and whilst this can damage the plant further, it also forms a cover over the leaf which prevents the aphids from moving or sucking more sap, and so causes starvation \citep{Prajneshu98}. The more aphids that have been on a leaf, the more honeydew there is and so the faster the aphids die, until the rate of death overtakes the rate of reproduction. \cite{Matis06} describe a model for the population growth of aphids with two species, the current population $\mathcal{X}_1$, and the cumulative population $\mathcal{X}_2$. It should be noted that although $\mathcal{X}_1$ and $\mathcal{X}_2$ are different `species' in terms of the reaction network, they are tracking the same population of aphids, but in different ways. The reaction list is
\begin{align*}
\begin{array}{rrrl}
\mathcal{R}_{1}: & \quad \mathcal{X}_1 &\xrightarrow{\phantom{a}c_{1}\phantom{a}} & 2\mathcal{X}_1 + \mathcal{X}_2\\
\mathcal{R}_{2}: & \quad \mathcal{X}_1 + \mathcal{X}_2 &\xrightarrow{\phantom{a}c_{2}\phantom{a}} & \mathcal{X}_2
\end{array}
\end{align*}

The CLE was taken as the inferential model, details of which can be found in Appendix~\ref{app:aphidlna}. Using parameter values inspired by the real data example in  \cite{whitaker2017a}, synthetic data were generated at 8 integer times using Gillespie's direct method with $c = (1.75, 0.001)'$ and ${X}_{1,0}={X}_{2,0}=5$. The data for $\{{X}_{2,t}\}$ were then discarded to obtain a challenging partial observation scenario, and the resulting data set was corrupted with Gaussian error. Following \cite{whitaker2017a}, the variance was taken as proportional to the current number of aphids in the system, which was found to give a better predictive fit in real data applications. Hence,
\begin{equation}\label{eqn:aphidObs}
Y_{t}=P' X_{t}+\varepsilon_{t},\qquad \varepsilon_{t}\sim \textrm{N}\left(0,\sigma^2 P' X_t \right),\qquad t=0,\ldots, 7
\end{equation}
where $P' = (1, 0)$.

\begin{figure}[ht]
\centering
\includegraphics[width=15cm,height=12cm]{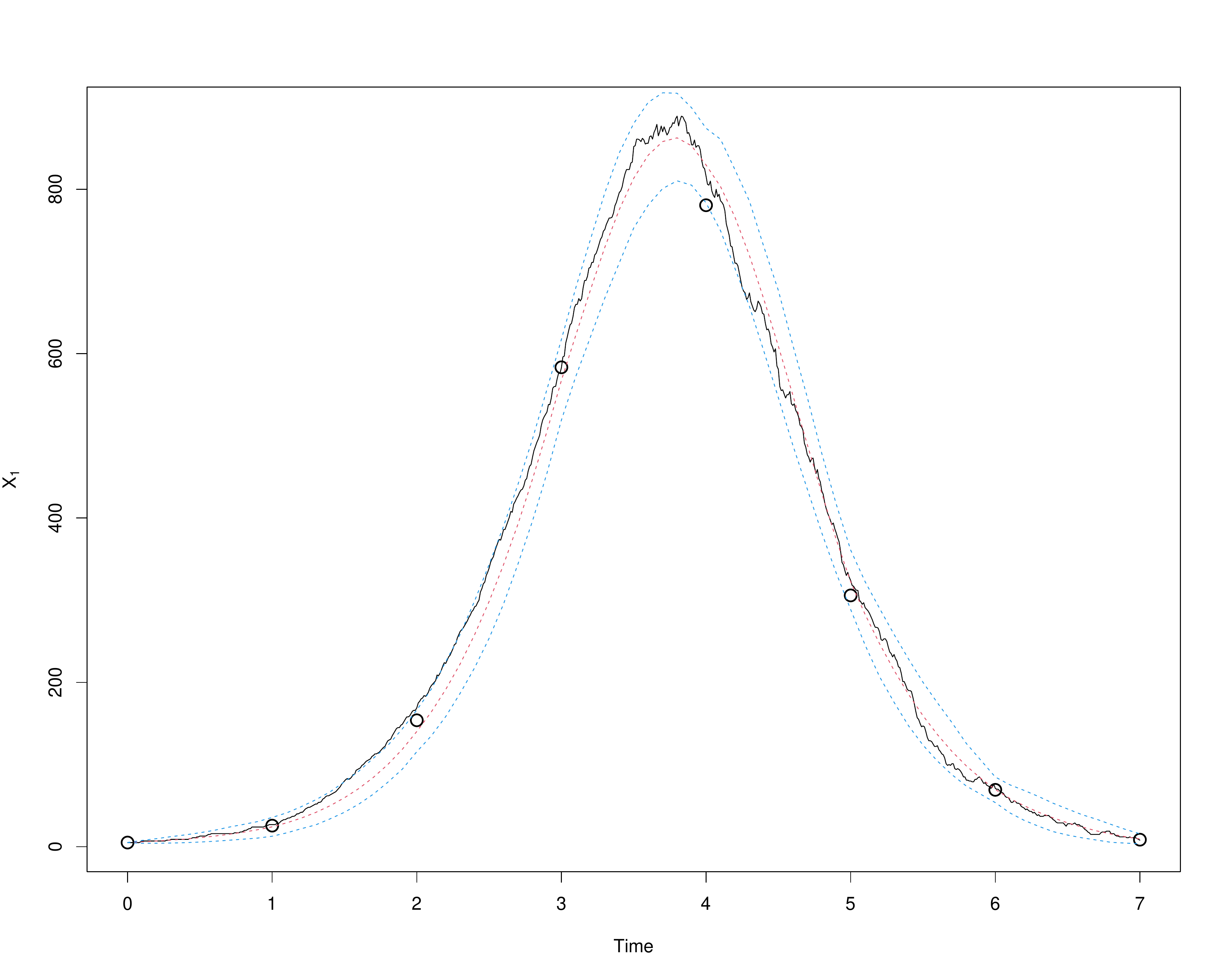}
\caption{Observations from the aphid data set, with the latent process (solid line) overlaid. The dashed lines are the mean, 2.5\% and 97.5\% quantiles of 1000 bridges generated using the ground truth for $c_1$ and $c_2$.}
\label{fig:latentProc3}
\end{figure}
The data are shown in Figure~\ref{fig:latentProc3}, alongside the underlying latent $X_{1,t}$ process that produced the data. It is clear that the behaviour of the latent process between observations is nonlinear. This precludes the use of bridge constructs that push the particles towards the observations in a linear fashion, such as the modified diffusion bridge of \cite{DurhGall02}.  A computationally inexpensive option is to use the myopic approach discussed in Section \ref{sec:bridge}. However, as previously mentioned, when the observation variance is small relative to the intrinsic stochasticity exhibited by the latent process, this implementation can lead to a highly variable estimator of the marginal likelihood, thus necessitating a far larger number of particles. This in turn can negate any computational benefit arising from the simplified form of the simulator and associated weight (compared to when using a bridge construct).

The performance of PMMH and CPMMH was therefore compared using either myopic simulation or the simple residual bridge. An independent prior specification was adopted, with $N(0, 10^2)$ distributions assigned to $\log c_1$ and $\log c_2$. The observation variance $\sigma$ and initial conditions ${X}_{1,0}$ and ${X}_{2,0}$ were treated as fixed and known. Using $\rho \approx 1$ in this application led to long term dependence between parameter draws, which reduced the effective sample size of the schemes. Therefore, $\rho$ was reduced to $0.75$ for this application, which was found to be optimal in terms of mESS of the resulting scheme. To implement the residual bridge construct, $\sigma^2 P' X_t$ was replaced by $\sigma^2 P'\eta_t$ where $\eta_t$ is the solution to (\ref{lna1}) at observation time $t$. It is emphasised that this is necessary to obtain a tractable bridge and does not introduce any further approximation in terms of the posterior output. 

\begin{figure}[ht]
\centering
\includegraphics[width=15cm,height=7.5cm]{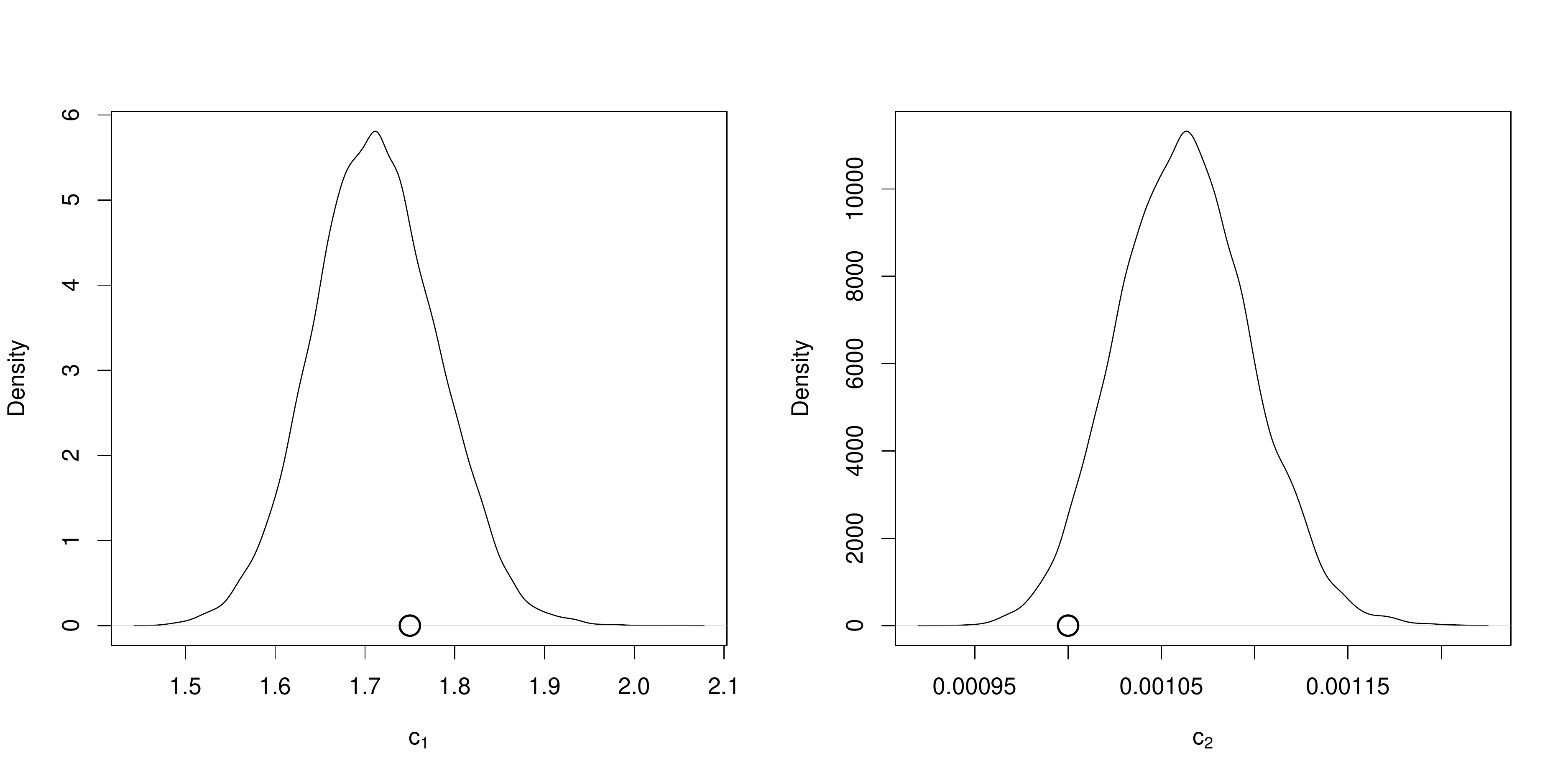}
\caption{Aphid model. Marginal posterior plots for the two parameters. The ground truth for each parameter is indicated by a circle on the corresponding plot.}
\label{fig:aphidPost}
\end{figure}

Figure~\ref{fig:aphidPost} and Table~\ref{tab:aphid} summarise the output of each scheme. It was found that using delayed acceptance and/or MALA gave no significant improvement in overall efficiency relative to the best performing schemes in Table~\ref{tab:aphid}. Relatively few particles are needed for the particle filter when using residual bridge constructs, and there are few observations in the dataset, consequently the computational gains provided by using a delayed acceptance step do not overcome the loss in statistical efficiency. Similarly, the partial observation regime limits the gradient information gained by MALA, meaning the additional cost of computing the gradient outweighs the statistical gains. Results for these schemes are therefore omitted. Table~\ref{tab:aphid} shows that the simple residual bridge-based schemes outperform the myopic schemes in terms of overall efficiency (by around a factor of 3). The behaviour of the simple residual bridge can be seen in Figure \ref{fig:latentProc3}, and adequately captures the dynamics of the latent process. Indeed, no improvement in overall efficiency when  using the residual bridge with additional subtraction was found (results omitted). Finally, note a small improvement in overall efficiency by solving the ODE system used by the residual bridge, once per iteration as opposed to once per particle. 

\begin{table}
\centering
\small
	\begin{tabular}{@{}lllllll@{}}
         \toprule
Scheme & $N$ & $\alpha$ & CPU (s) & mESS & mESS/s & Rel.  \\
\midrule
PMMH / RWM / Myopic & 100 & 0.10 & 15320 & 4172 & 0.272 & 1.0 \\ 
CPMMH / RWM / Myopic & \phantom{0}35 & 0.09 & \phantom{0}4452 & 2482 & 0.558 & 2.1 \\ 
\midrule
PMMH / RWM / $\textrm{RB}_\textrm{part}$ & \phantom{00}5 & 0.19 & \phantom{0}6527 & 4857 & 0.744 & 2.7 \\
PMMH / RWM / $\textrm{RB}_\textrm{iter}$ & \phantom{00}5 & 0.17 & \phantom{0}5030 & 4226 & 0.840 & 3.1 \\
\midrule
CPMMH / RWM / $\textrm{RB}_\textrm{part}$ & \phantom{00}2 & 0.19 & \phantom{0}2593 & 3563   & 1.374  & 5.1 \\ 
CPMMH / RWM / $\textrm{RB}_\textrm{iter}$ & \phantom{00}2 & 0.18 & \phantom{0}2493 & 3737   & 1.499  & 5.5 \\ 
\bottomrule
\end{tabular}
      \caption{Aphid model. Number of particles $N$, acceptance rate $\alpha$, CPU time (in seconds), minimum ESS, minimum ESS per second, and relative (to the worst performing scheme) minimum ESS per second. All results are based on $10^5$ iterations of each scheme.}\label{tab:aphid}
\end{table}

\subsection{Lotka-Volterra} \label{sec:LV}

The Lotka-Volterra system consists of two species, prey ($\mathcal{X}_1$) and predator ($\mathcal{X}_2$), and three reactions: $\mathcal{R}_1$ denotes the reproduction of a member of the prey species, $\mathcal{R}_2$ denotes the death of a member of prey and the reproduction of a predator, and $\mathcal{R}_3$ denotes the death of a predator. The resulting reaction list is
\begin{align*}
\begin{array}{rrrl}
\mathcal{R}_1: & \quad \mathcal{X}_1 &\xrightarrow{\phantom{a}c_1\phantom{a}} & 2\mathcal{X}_1\\
\mathcal{R}_2: & \quad \mathcal{X}_1 + \mathcal{X}_2 &\xrightarrow{\phantom{a}c_2\phantom{a}} & 2\mathcal{X}_2\\
\mathcal{R}_3: & \quad \mathcal{X}_2 &\xrightarrow{\phantom{a}c_3\phantom{a}} & \emptyset\\
\end{array}
\end{align*}    
The system is typically used to benchmark competing inference algorithms; see e.g. \cite{BWK08}, \cite{koblents2015} when using the MJP representation or \cite{Fuchs_2013}, \cite{ryder21}, \cite{graham17}, \cite{GoliBrad19} when using the CLE. Here, we adopt the latter as the inferential model (see Appendix~\ref{app:lvlna} for details of the CLE derivation).

\begin{figure}[ht]
\centering
\includegraphics[width=15cm,height=5cm]{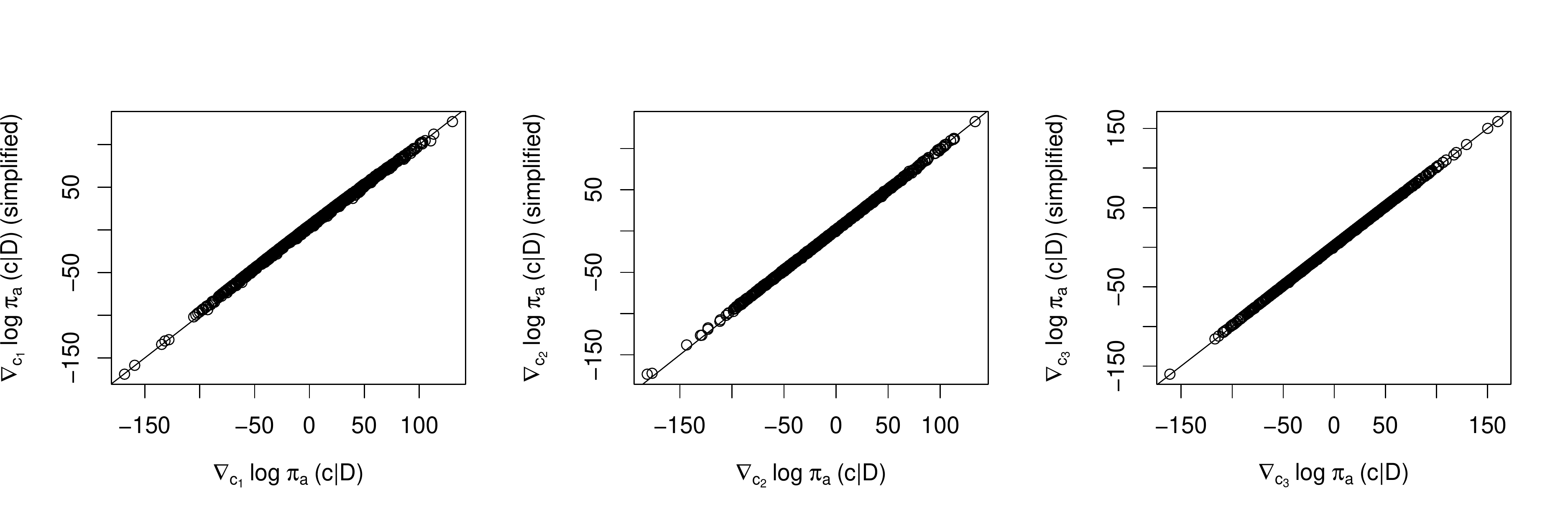}
\caption{Lotka-Volterra model. Full versus simplified gradient of the log posterior density with respect to $c_1$ (left), $c_2$ (centre) and $c_3$ (right) computed for $1000$ draws from the joint posterior over $c$.}
\label{fig:lvMALA}
\end{figure}

A single realisation of the jump process at 51 integer times was generated via Gillespie's direct method 
with rate constants as in \cite{BWK08}, that is $c=(0.5,0.0025,0.3)'$ and an initial condition of $X_{0}=(100,100)'$. The data for both species were corrupted with independent, additive Gaussian error and standard deviation $\sigma=1$. The corresponding observation equation (\ref{obs_eq}) becomes
\[
Y_{t}=X_{t}+\varepsilon_{t},\qquad \varepsilon_{t}\sim \textrm{N}\left(0,\operatorname{diag}(\sigma^2,\sigma^2)\right),\qquad t=0,\ldots, 50.
\]
An independent prior specification for $c$ was assumed, with $N(0,10^2)$ distributions assigned to the logarithm of each rate constant. CPMMH was run with 4 different bridge implementations:  $\textrm{RB}_\textrm{iter}$, $\textrm{RB}_\textrm{part}$, $\textrm{RB}_\textrm{iter}^-$, and $\textrm{RB}_\textrm{part}^-$, along with the presence or absence of two techniques: simplified MALA and delayed acceptance. 
Since the residual bridge with extra subtraction ($\textrm{RB}^-$) typically outperformed the simple residual bridge ($\textrm{RB}$) up to a factor of 2 in terms of overall efficiency (depending on the acceleration technique employed), results are reported for $\textrm{RB}^-$ only. Similarly, little difference between the gradients employed by simplified MALA versus full MALA (see Figure~\ref{fig:lvMALA}) was found; results for the former are reported.  

\begin{figure}[ht]
\centering
\includegraphics[width=15cm,height=5cm]{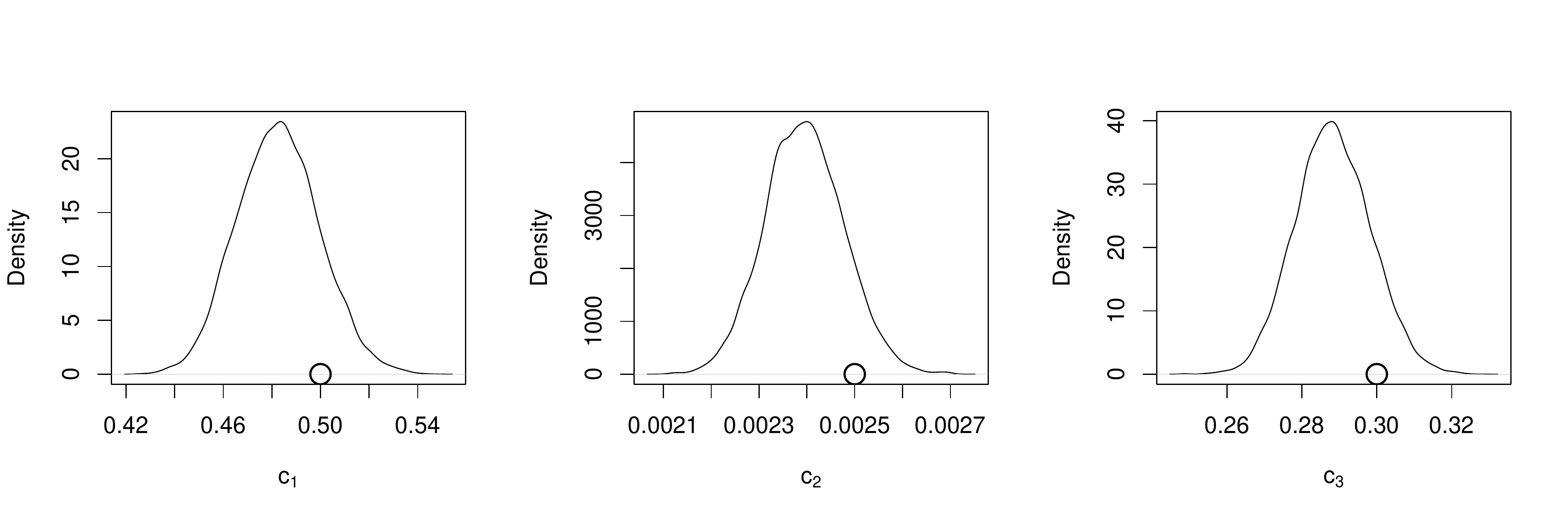}
\caption{Lotka-Volterra model. Marginal posterior plots for the three parameters. The ground truth for each parameter is indicated by a circle on the corresponding plot.}
\label{fig:lvPost}
\end{figure}

All schemes were implemented for $10^5$ iterations, including PMMH (with $\textrm{RB}_\textrm{iter}^-$ which performed best of all bridge implementations) for benchmarking. Figure~\ref{fig:lvPost} and Table~\ref{tab:LV} summarise findings. The former gives marginal parameter posterior densities from the output of the best performing inference scheme (with consistent results obtained from other schemes but not shown) from which we see consistency with the ground truth values. From Table~\ref{tab:LV}, the most basic CPMMH scheme (without MALA or delayed acceptance) gives an improvement in overall efficiency over PMMH of around a factor of 3. It is also clear that while the \emph{per iteration} implementation of $\textrm{RB}^-$ results in a small reduction in minimum effective sample size compared to the \emph{per particle} implementation, the computational saving is worthwhile. Replacing the RWM parameter proposal with MALA gives a relative increase in overall efficiency by a factor of 3. It is evident that the combination of delayed acceptance and MALA gives the best performing scheme, with an order of magnitude increase in mESS/s against the benchmark.         

\begin{table}
\centering
\small
	\begin{tabular}{@{}lllllllll@{}}
         \toprule
Scheme & $N$ & $\alpha_{1}$ & $\alpha_{2|1}$ & $\alpha$ & CPU (s) & mESS & mESS/s & Rel.  \\
\midrule
PMMH / RWM / $\textrm{RB}_\textrm{iter}^-$ & 2 & -- & -- & 0.13 & 25735 & \phantom{0}2809 & 0.109 & \phantom{0}1.0 \\
\midrule
CPMMH / RWM / $\textrm{RB}_\textrm{iter}^-$ & 2 & -- & -- & 0.21 & 25372 & \phantom{0}7705 & 0.304 & \phantom{0}2.8  \\
CPMMH / RWM / $\textrm{RB}_\textrm{part}^-$ & 2 & -- & -- & 0.25 & 31568 & \phantom{0}8445 & 0.280 & \phantom{0}2.6 \\
\midrule
CPMMH / sMALA / $\textrm{RB}_\textrm{iter}^-$ & 2 & -- & -- & 0.44 & 28898 & 24709 & 0.855 & \phantom{0}7.8 \\
CPMMH / sMALA / $\textrm{RB}_\textrm{part}^-$ & 2 & -- & -- & 0.44 & 39643 & 25545 & 0.644 & \phantom{0}5.9 \\
\midrule
daCPMMH / RWM / $\textrm{RB}_\textrm{iter}^-$ & 2 & 0.22 & 0.85 & 0.19 & 10877 & \phantom{0}6415 & 0.590 & \phantom{0}5.4  \\
daCPMMH / sMALA / $\textrm{RB}_\textrm{iter}^-$ & 2 & 0.46 & 0.84 & 0.39 & 18339 & 19944 & 1.088 & 10.0  \\
\bottomrule
\end{tabular}
      \caption{Lotka-Volterra model. Number of particles $N$, acceptance rates $\alpha_1$, $\alpha_{2|1}$ and $\alpha$, CPU time (in seconds), minimum ESS, minimum ESS per second, and relative (to the worst performing scheme) minimum ESS per second. All results are based on $10^5$ iterations of each scheme.}\label{tab:LV}
\end{table}

\section{Discussion}
\label{sec:disc}

This work considered the problem of Bayesian inference for the parameters governing two commonly used representations of a stochastic kinetic model (SKM), namely the Markov jump process (MJP) and (time-discretised) chemical Langevin equation (CLE) representations. Although the MJP is the most natural description for the dynamics of a collection of species whose discrete-valued states vary continuously in time, it is often eschewed in favour of the CLE which models species dynamics according to a nonlinear multivariate It\^o diffusion process. Inference under either approach is made challenging by the intractability of the observed data likelihood, whose computation requires integrating over reaction times and types (in the case of the MJP) or intermediate states (in the case of the CLE) between observation instants (although for the CLE, the number of intermediate states is controlled by the practitioner, lending to its appeal). Consequently, the usual method of choice is particle Markov chain Monte Carlo (particle MCMC) whereby a bootstrap particle filter is used to give realisations of an unbiased estimator of the intractable likelihood. For systems where this approach is likely to be computationally prohibitive, an inexpensive surrogate can be used as the inferential model. Work in this direction has included, \emph{inter alia}, Gaussian process emulation \citep[see e.g.][]{rasmussen03,fielding11} and direct approximation of the SKM using the linear noise approximation \cite[LNA, see e.g.][]{Komorowski09,stathopoulos13,fearnhead14}. 

Recent work that has attempted to leverage the tractability of the LNA to accelerate inference under the SKM (rather than the surrogate) includes the use of delayed acceptance \citep{Goli15} and bridge constructs \citep{whitaker2017,Goli19}. The contribution here is a novel combination of these techniques, to which a gradient-based parameter proposal via the Metropolis adjusted Langevin algorithm (MALA) and the use of correlated particle filters are added. The result is a unified inference framework that targets the posterior under either the MJP or CLE. Two implementations of the surrogate-based bridge construct were considered; one in which the ODE system governing the LNA is re-solved per particle (with initial conditions informed by the current state particle), and one in which the ODE system is solved once per iteration. For the two applications that allowed this comparison, it was found that a small increase in overall efficiency was possible by solving the ODE system once per iteration. Not surprisingly, the likely difference depends on the number of particles employed by the particle filter and on the type of bridge being implemented; for example, the residual bridge requires the integration of order $s$ components versus order $s^2$ when using the extra subtraction. Nevertheless, use of a bridge construct (irrespective of the aforementioned implementation options) is crucial in obtaining significant improvements in overall efficiency compared to standard (C)PMMH implementations based on forward simulation.  

In addition to the use of LNA-based bridge constructs, the use of the gradient of the log posterior under the LNA to obtain a MALA proposal, and the use of the LNA likelihood in a delayed acceptance scheme were investigated. The results suggest that these approaches are likely to be most effective in scenarios where the overall computational cost is dominated by the particle filter; that is, when obtaining the LNA-based gradient and likelihood is relatively inexpensive. It is well known \citep[see e.g.][]{roberts98} that MALA performs increasingly well in terms of statistical efficiency compared to random walk Metropolis (RWM) as the dimension of the target increases; nevertheless, we were able to demonstrate an order of magnitude increase in overall efficiency for a 3-dimensional target. It is anticipated that greater improvements in relative mixing efficiency are possible as the number of rate constants $r$ increases, but note that this must be tempered by the increased computational cost associated with solving the associated ODE system, which scales linearly in $r$. Study of the conditions under which greater improvements in efficiency can be made merits further attention.

In scenarios when only a few (e.g. $N<10$) particles are required, use of a simplified gradient requiring fewer ODE components to be integrated was investigated. The success of this approach depends on the extent to which the dependence of the LNA variance on the parameters can be ignored and merits further consideration. Use of the surrogate gradient inside a Hamiltonian Monte Carlo \citep[HMC,][]{duane87} scheme is also of interest to us. HMC was successfully applied in the SKM setting by \cite{stathopoulos13}, albeit in a sacrificial observation setting, and with the LNA as the inferential model. It is expected that in the setting considered here, HMC would not perform well when using the simplified surrogate gradient, as ignoring the dependence of the variance on the parameters means that potential energy will not be conserved even if the step size used in the numerical integrator is very small, leading to lower overall acceptance rates. Nevertheless, further work is required to assess the potential benefits of using HMC with the full surrogate gradient, for observation scenarios such as those considered here.



\bibliographystyle{apalike}
\bibliography{bridgebib}

\appendix

\section{Validity of delayed acceptance CPMMH} \label{valid_app}

It is shown that a delayed acceptance Metropolis-Hastings scheme with proposal kernel $q(c^*|c)K(u^*|u)$ and the acceptance probability given in (\ref{overall_acc_prob}) satisfies detailed balance with respect to the target density $\pi(c)\hat{p}_u(\mathcal{D}|c)g(u)$. Moves that are rejected satisfy detailed balance trivially, as the new state of the chain is equal to the previous state. When the chain does move, we have that
\begin{align*}
    &\pi(c)\hat{p}_u(\mathcal{D}|c)g(u)q(c^*|c)K(u^*|u)\alpha_1\left(c^*|c\right)\alpha_{2|1}\left\{(c^*,u^*)|(c,u)\right\} \\
    &= \pi(c)p_\LNA(\mathcal{D}|c)q(c^*|c)\alpha_1\left(c^*|c\right)\times \frac{\hat{p}_u(\mathcal{D}|c)g(u)K(u^*|u) }{p_\LNA(\mathcal{D}|c)}\alpha_{2|1}\left\{(c^*,u^*)|(c,u)\right\}.
\end{align*}
Now, 
\begin{align*}
\pi(c)p_\LNA(\mathcal{D}|c)q(c^*|c)\alpha_1\left(c^*|c\right)&=\textrm{min} \left\{ \pi(c)p_\LNA(\mathcal{D}|c)q(c^*|c), \pi(c^*) p_\LNA(\mathcal{D}|c^*)q(c|c^*) \right\}
\end{align*}
which is clearly symmetric in $(c,u)$ and $(c^*,u^*)$. Similarly, 
\begin{align*}
&\frac{\hat{p}_u(\mathcal{D}|c)g(u)K(u^*|u) }{p_\LNA(\mathcal{D}|c)}\alpha_{2|1}\left\{(c^*,u^*)|(c,u)\right\} \\
&=  \textrm{min} \left\{ \frac{\hat{p}_u(\mathcal{D}|c)g(u)K(u^*|u) }{p_\LNA(\mathcal{D}|c)}, \frac{\hat{p}_{u^*}(\mathcal{D}|c^*)g(u^*)K(u|u^*)}{p_\LNA(\mathcal{D}|c^*)} \right\}   
\end{align*}
since $g(u)K(u^*|u) = g(u^*)K(u|u^*)$ from (\ref{gK_db}), and we again have symmetry in $(c,u)$ and $(c^*,u^*)$. Hence, detailed balance is satisfied and the daCPMMH schemes targets $\pi(c)\hat{p}_u(\mathcal{D}|c)g(u)$ for which $\pi(c|\mathcal{D})$ is a marginal density.

\section{Residual bridge details}
\label{app:rbridge}

Without loss of generality, consider a time interval $[0,T]$ partitioned as
\begin{equation}
0 = \tau_0 < \tau_1 < \ldots < \tau_{m-1} < \tau_m = T,
\label{partition}
\end{equation}
with $\tau_{i+1} - \tau_i = \Delta\tau = T/m$. Suppose that $x_0$ and $y_T$ are observed at times $0$ and $T$, and we seek a density $q(x_{(0,T]}|x_{0},y_{T},c)$ corresponding to either the residual bridge (RB) or residual bridge with additional subtraction (RB$^-$). 

For the residual bridge, we partition $X_t$ as $X_t=\eta_t+R_t$ where $\eta_t$ satisfies
\[
\frac{d\eta_t}{dt} = Sh(\eta_t).
\]
By partitioning $X_t$ as above, (\ref{obs_eq}) can be written as
\[
Y_T - P'\eta_T = P'R_T + \epsilon_T, \quad \epsilon_T \sim N(0, \Sigma).
\]
Then, the modified diffusion bridge can be used to approximate the joint distribution of $R_{\tau_{k+1}}$ and $Y_T - P'\eta_T$ conditional on the residual process at the previous time point $r_{\tau_k}$ to obtain
\[
\left.
\left(
\begin{pmatrix}
R_{\tau_{k+1}} \\
Y_T - P'\eta_T 
\end{pmatrix}
\right|
r_{\tau_k} \right) \sim N \left\{
\begin{pmatrix}
r_{\tau_k} + (\alpha_k-\delta_k^\eta)\Delta\tau \\
P'\{\eta_T + r_{\tau_k} + (\alpha_k-\delta_k^\eta)\Delta_k\}
\end{pmatrix},
\begin{pmatrix}
\beta_k\Delta\tau & \beta_k P \Delta\tau \\
P'\beta_k \Delta\tau & P'\beta_k P \Delta_k + \Sigma
\end{pmatrix}
\right\}.
\]
Here, $\alpha_k = Sh(x_{\tau_k})$, $\beta_k = S\operatorname{diag}\{h(x_{\tau_k})\}S'$, $\Delta_k = T - \tau_k$, and $\delta_k^\eta$ is an approximation of $d\eta/dt$ give by
\[
\delta_k^\eta = \frac{\eta_{\tau_{k+1}}-\eta_{\tau_k}}{\Delta\tau}.
\]
Conditioning on $y_T - P'\eta_T$ gives
\[
\left(R_{\tau_{k+1}}|r_{\tau_k}, y_T \right) \sim N(\mu_{\textrm{RB}}, \Psi_{\textrm{RB}}),
\]
where
\begin{equation}
\mu_{\textrm{RB}} = r_{\tau_k} + (\alpha_k-\delta_k^\eta)\Delta\tau + \beta_k P \Delta\tau (P'\beta_k P\Delta_k + \Sigma)^{-1} [y_T - P'\{\eta_T+r_{\tau_k} + (\alpha_k-\delta_k^\eta)\Delta_k\}],
\label{eq:RBmean}
\end{equation}
and
\begin{equation}
\Psi_{\textrm{RB}} = \beta_k\Delta\tau - \beta_k P \Delta\tau(P'\beta_k P\Delta_k + \Sigma)^{-1} P'\beta_k \Delta \tau. 
\label{eq:RBvar}
\end{equation}
The partition of $X_t$ then yields
\[
\left(X_{\tau_{k+1}}|x_{\tau_k}, y_T \right) \sim N(\eta_{\tau_{k+1}} +  \mu_{\textrm{RB}}, \Psi_{\textrm{RB}}).
\]

If $\eta_t$ doesn't adequately capture the dynamics of the target process, a second residual bridge can be obtained by instead partitioning $X_t$ as $X_t = \eta_t + \hat{\rho}_t + R_t^-$, where $\hat{\rho}_t = \operatorname{E}(\hat{R}_t|r_0, y_T$) is an approximation of the conditional expected value of the residual process, and $R_t^-$ is the residual stochastic process that now remains after this further decomposition. We obtain $\hat{R}_t$ using the LNA; see Section \ref{sec:lna}. Thus, the joint distribution of $\hat{R}_t$ and $Y_T - P'\eta_T$ (conditional on $\hat{r}_0$) is
\[
\left.
\left(
\begin{pmatrix}
\hat{R}_t \\
Y_T - P'\eta_T 
\end{pmatrix}
\right|
\hat{r}_0 \right) \sim N \left\{
\begin{pmatrix}
G_t\hat{r}_0 \\
P'G_T \hat{r}_0
\end{pmatrix},
\begin{pmatrix}
V_t & V_t(G_t')^{-1}G_T'P \\
P'G_TG_t^{-1}V_t & P'V_T P + \Sigma
\end{pmatrix}
\right\}.
\]
Conditioning on $y_T-P'\eta_T$ gives 
\[
\hat{\rho}_t = \operatorname{E}[\hat{R}_t|r_0,y_T] = G_t\hat{r}_0 + V_t(G_t')^{-1}G_T P(P'V_TP + \Sigma)^{-1}(y_T - P'\eta_T - P'G_T \hat{r}_0).
\]
The modified diffusion bridge for the residual process (after subtracting both the drift and our conditional expected residual term) is then constructed by finding the approximate joint distribution of $R_{\tau_{k+1}}^-$ and $Y_T-P'(\eta_T+\hat{\rho}_T)$ conditional on $r_{\tau_k}^-$. This step follows as above and so it is omitted for brevity. Conditioning on $y_T - P'(\eta_T+\hat{\rho}_T)$ gives 
\[
\left(R_{\tau_{k+1}}^-|r_{\tau_k}^-, y_T \right) \sim N(\mu_{\textrm{RB}^-}, \Psi_{\textrm{RB}^-}),
\]
where $\Psi_{\textrm{RB}^-} = \Psi_{\textrm{RB}}$, and
\begin{equation}
\mu_{\textrm{RB}^-} = r_{\tau_k} + (\alpha_k-\delta_k^\eta-\delta_k^\rho)\Delta\tau + \beta_k P \Delta\tau (P'\beta_k P\Delta_k + \Sigma)^{-1} [y_T - P'\{\eta_T + \hat{\rho}_T + r_{\tau_k}^- + (\alpha_k-\delta_k^\eta-\delta_k^\rho)\Delta_k\}].
\label{eq:RBminusmean}
\end{equation}
Here,
\[
\delta_k^\rho = \frac{\hat{\rho}_{\tau_{k+1}}-\hat{\rho}_{\tau_k}}{\Delta\tau}.
\]
Finally, use the partition of $X_t$ to give 
\[
\left(X_{\tau_{k+1}}|x_{\tau_k}, y_T \right) \sim N(\eta_{\tau_{k+1}} + \hat{\rho}_{\tau_{k+1}}+\mu_{\textrm{RB}^-}, \Psi_{\textrm{RB}^-}).
\]

\section{Accelerated PMMH: general algorithm}\label{app:general}
Here, an overview (see Algorithm~\ref{algcor}) of the pseudo-marginal Metropolis-Hastings algorithm augmented by the acceleration techniques described in Section~\ref{sec:acc} is presented. In the applications considered in Section~\ref{sec:apps}, a fixed and known initial condition $x_{t_0}$ is assumed; Algorithm~\ref{auxPF} is therefore initialised with equally weighted particles $x_{t_0}^{(k)}=x_{t_0}$ for all $k$, and Algorithm~\ref{algLNAff} is initialised with $a_{t_0}=x_{t_0}$ and $B_{t_0}=0$. For unknown $x_{t_0}$, the parameter vector $c$ can be augmented to include the components of $x_{t_0}$, and the initialisation choices follow as above.      

\begin{algorithm}[h!]
\caption{Accelerated PMMH}\label{algcor}
\textbf{Input:} parameter proposal tuning parameters $\Sigma_T$ and $\lambda$, correlation parameter $\rho$, number of particles $N$ and the number of iterations $n_{\textrm{iters}}$.
\begin{enumerate}
\item \textbf{For iteration $j=0$:}
\begin{itemize}
\item[(a)] Set $c^{(0)}$ in the support of $\pi(c)$ and initialise the auxiliary variable $u^{(0)}\sim \textrm{N}(0,I_d)$.
\item[(b)] Compute ${p}_\LNA(\mathcal{D}|c^{(0)})$ and $\nabla \log p_\LNA(\mathcal{D}|c^{(0)})$ via recursive application of Algorithm~\ref{algLNAff}. Hence compute $\nabla \log \pi_\LNA(c^{(0)}|\mathcal{D})$.  
\item[(c)] Compute $\hat{p}_{u^{(0)}}(\mathcal{D}|c^{(0)})$ via recursive application of Algorithm~\ref{auxPF}.
\end{itemize}
\item \textbf{For iteration $i=1,\ldots, n_{\textrm{iters}}$:}
\begin{itemize}
\item[(a)] Draw $Z \sim N(0,\Sigma_T)$ and set $c^* = c^{(j-1)} + \frac{\lambda^2}{2}\Sigma_T \nabla \log \left(\pi_\LNA(c^{(j-1)}|\mathcal{D})\right) + \lambda Z$.
\item[(b)] \textbf{Stage One:}
\begin{itemize}
\item[(i)] Compute ${p}_\LNA(\mathcal{D}|c^*)$ and $\nabla \log p_\LNA(\mathcal{D}|c^*)$ via recursive application of Algorithm~\ref{algLNAff}. Hence compute $\nabla \log \pi_\LNA(c^*|\mathcal{D})$.
\item[(ii)] With probability $\alpha_1(c^*|c)$ as given by (\ref{stage1}), 
go to 2(c), otherwise set $c^{(j)}=c^{(j-1)}$, $u^{(j)}=u^{(j-1)}$, ${p}_\LNA(\mathcal{D}|c^{(j)})={p}_\LNA(\mathcal{D}|c^{(j-1)})$, $\nabla \log \pi_\LNA(c^{(j)}|\mathcal{D})=\nabla \log \pi_\LNA(c^{(j-1)}|\mathcal{D})$, $\hat{p}_{u^{(j)}}(\mathcal{D}|c^{(j)})=\hat{p}_{u^{(j-1)}}(\mathcal{D}|c^{(j-1)})$, increment $j$ and return to step 2(a). 
\end{itemize}
\item[(c)] \textbf{Stage Two:}
\begin{itemize}
\item[(i)] Draw $\omega\sim \textrm{N}(0,I_d)$. Put $u^*=\rho u^{(j-1)}+\sqrt{1-\rho^2}\omega$.
\item[(ii)]  Compute $\hat{p}_{u^*}(\mathcal{D}|c^*)$ via recursive application of Algorithm~\ref{auxPF}.
\item[(iii)] With probability $\alpha_{2|1}\{(c^*,u^*)|(c^{(j-1)},u^{(j-1)})\}$ as given by (\ref{stage2}), put $c^{(j)}=c^*$, $u^{(j)}=u^*$, ${p}_\LNA(\mathcal{D}|c^{(j)})={p}_\LNA(\mathcal{D}|c^*)$, $\nabla \log \pi_\LNA(c^{(j)}|\mathcal{D})=\nabla \log \pi_\LNA(c^*|\mathcal{D})$, $\hat{p}_{u^{(j)}}(\mathcal{D}|c^{(j)})=\hat{p}_{u^*}(\mathcal{D}|c^*)$. Otherwise set $c^{(j)}=c^{(j-1)}$, $u^{(j)}=u^{(j-1)}$, ${p}_\LNA(\mathcal{D}|c^{(j)})={p}_\LNA(\mathcal{D}|c^{(j-1)})$, $\nabla \log \pi_\LNA(c^{(j)}|\mathcal{D})=\nabla \log \pi_\LNA(c^{(j-1)}|\mathcal{D})$, $\hat{p}_{u^{(j)}}(\mathcal{D}|c^{(j)})=\hat{p}_{u^{(j-1)}}(\mathcal{D}|c^{(j-1)})$. Increment $j$ and return to step 2(a). 
\end{itemize}
\end{itemize}
\end{enumerate}
\textbf{Output:} $c^{(1)},\ldots,c^{(n_{\textrm{iters}})}$.
\end{algorithm}

\section{Applications: further modelling details\label{app:models}}

\subsection{Aphid growth model} \label{app:aphidlna}
Let $X_t=(X_{1,t},X_{2,t})'$ denote the state of the system at time $t$. The stoichiometry matrix associated with the reaction system is
$$
S = \left(\begin{array}{rr} 
1 & -1\\
 1 & 0
\end{array}\right)
$$
and the associated hazard function is
$$
h(x_t,c) = (c_1 x_{1,t}, c_2 x_{1,t} x_{2,t})'.
$$
The CLE for this model is
\begin{align}\label{eq:aphidcle}
d\begin{pmatrix} X_{1,t}\\X_{2,t}\end{pmatrix}
&=\begin{pmatrix}
				c_{1}x_{1,t} - c_2 x_{1,t} x_{2,t} \\
				c_{1}x_{1,t}
			\end{pmatrix}dt +
\begin{pmatrix}
			c_{1}x_{1,t} + c_2 x_{1,t} x_{2,t} & c_{1}x_{1,t} \\
			 c_{1}x_{1,t}	 & c_{1}x_{1,t}
			\end{pmatrix}^{1/2}d \begin{pmatrix} W_{1,t}\\W_{2,t}\end{pmatrix}. \nonumber
\end{align}
Similarly, the LNA for this model is specified by the coupled ODE system
\begin{align*}
\frac{d\eta_t}{dt} &= (c_1 \eta_{1,t} - c_2 \eta_{1,t}\eta_{2,t}, c_1 \eta_{1,t})', \\
\frac{dG_t}{dt} &= 
\begin{pmatrix}
c_1 - c_2 \eta_{2,t} & -c_2 \eta_{1,t} \\
c_1 & 0
\end{pmatrix}
G_t \\
\frac{dV_t}{dt} &= V_t
\begin{pmatrix}
c_1 - c_2 \eta_{2,t} & c_1 \\
-c_2 \eta_{1,t} & 0
\end{pmatrix}
+
\begin{pmatrix}
c_1 \eta_{1,t} + c_2 \eta_{1,t}\eta_{2,t} & c_1 \eta_{1,t} \\
c_1 \eta_{1,t} & c_1 \eta_{1,t}
\end{pmatrix}
+
\begin{pmatrix}
c_1 - c_2 \eta_{C,t} & -c_2 \eta_{N,t} \\
c_1 & 0
\end{pmatrix}
V_t.
\end{align*} 

\subsection{Lotka-Volterra model} 
\label{app:lvlna}
Let $X_{t}=(X_{1,t},X_{2,t})'$ denote the system state at time $t$. 
The stoichiometry matrix associated with the reaction system is given by
\[
S = \left(\begin{array}{rrr} 
1 & -1 & 0 \\
0 & 1 & -1 
\end{array}\right)
\]
and the associated hazard function is 
\[
h(x_{t},c) = (c_1 x_{1,t},c_2 x_{1,t}x_{2,t},c_3 x_{2,t})'.
\] 
The CLE for this model is given by
\[
d\begin{pmatrix}X_{1,t} \\X_{2,t} \end{pmatrix}=
\begin{pmatrix}c_1 x_{1,t}-c_2 x_{1,t}x_{2,t} \\c_2 x_{1,t}x_{2,t}-c_3 x_{2,t} \end{pmatrix}\,dt+
\begin{pmatrix}c_1 x_{1,t}+c_2 x_{1,t}x_{2,t}   & -c_2 x_{1,t}x_{2,t} \\ -c_2 x_{1,t}x_{2,t} &c_2 x_{1,t}x_{2,t}+c_3 x_{2,t} 
\end{pmatrix}^{\frac{1}{2}}\,d\begin{pmatrix}W_{1,t} \\W_{2,t} \end{pmatrix}
\]
where $W_{1,t}$ and $W_{2,t}$ are independent standard Brownian motion processes. The LNA for this model is specified by the coupled ODE system
\begin{align*}
\frac{d\eta_t}{dt} &= (c_1 \eta_{1,t} - c_2 \eta_{1,t}\eta_{2,t}, c_2 \eta_{1,t}\eta_{2,t} - c_3 \eta_{2,t})', \\
\frac{dG_t}{dt} &= 
\begin{pmatrix}
c_1 - c_2 \eta_{2,t} & -c_2 \eta_{1,t} \\
c_2 \eta_{2,t} & c_2 \eta_{1,t} - c_3
\end{pmatrix}
G_t \\
\frac{dV_t}{dt} &= V_t
\begin{pmatrix}
c_1 - c_2 \eta_{2,t} & c_2 \eta_{2,t} \\
-c_2 \eta_{1,t} & c_2 \eta_{1,t} - c_3
\end{pmatrix}
+
\begin{pmatrix}
c_1 \eta_{1,t} + c_2 \eta_{1,t}\eta_{2,t} & -c_2 \eta_{1,t}\eta_{2,t} \\
-c_2 \eta_{1,t}\eta_{2,t} & c_2 \eta_{1,t}\eta_{2,t} + c_3 \eta_{2,t}
\end{pmatrix}\\
&\phantom{=} +
\begin{pmatrix}
c_1 - c_2 \eta_{2,t} & -c_2 \eta_{1,t} \\
c_2 \eta_{2,t} & c_2 \eta_{1,t} - c_3
\end{pmatrix}
V_t.
\end{align*} 

\end{document}